\documentclass[fleqn,usenatbib]{mnras}

\usepackage{newtxtext,newtxmath}
\usepackage[T1]{fontenc}

\DeclareRobustCommand{\VAN}[3]{#2}
\let\VANthebibliography\thebibliography
\def\thebibliography{\DeclareRobustCommand{\VAN}[3]{##3}\VANthebibliography}

\usepackage{graphicx}	% Including figure files
\usepackage{amsmath}	% Advanced maths commands
\usepackage[dvipsnames]{xcolor}
\usepackage{stfloats}

\newcommand{\hMpc}{h^{-1}\,\mathrm{Mpc}}
\newcommand{\Mpc}{\mathrm{Mpc}}
\newcommand{\hinvMpc}{h\,\mathrm{Mpc}^{-1}}
\newcommand{\invMpc}{\mathrm{Mpc}^{-1}}
\newcommand{\emu}{\texttt{COMET}}
\newcommand{\Euclid}{\emph{Euclid}}
\newcommand{\bx}{\boldsymbol{x}}
\newcommand{\bk}{\boldsymbol{k}}
\newcommand{\bs}{\boldsymbol{s}}
\newcommand{\bu}{\boldsymbol{u}}
\newcommand{\bv}{\boldsymbol{v}}
\newcommand{\bq}{\boldsymbol{q}}

\newcommand{\beq}{\begin{equation}}
\newcommand{\eeq}{\end{equation}}
\newcommand{\beqa}{\begin{eqnarray}}
\newcommand{\eeqa}{\end{eqnarray}}

\newcommand{\Plin}{P_{\rm{L}}}
\newcommand{\kv}{\boldsymbol{k}}
\newcommand{\qv}{\boldsymbol{q}}

\newcommand{\Pnw}{P_{\rm{nw}}}
\newcommand{\Pw}{P_{\rm{w}}}
\newcommand{\Peh}{P_{\rm{EH}}}
\newcommand{\ks}{k_{\rm{s}}}
\newcommand{\kosc}{k_{\rm{osc}}}

\title[\texttt{COMET}: Emulating perturbation theory]{\texttt{COMET}: Clustering Observables Modelled by Emulated perturbation Theory}

\author[A. Eggemeier et al.]{
Alexander Eggemeier,$^{1}$\thanks{E-mail: aeggemeier@astro.uni-bonn.de}\thanks{Argelander Fellow}
Benjamin Camacho-Quevedo,$^{2,3}$
Andrea Pezzotta,$^{4}$
Martin Crocce,$^{2,3}$
\newauthor
Rom\'an Scoccimarro,$^{5}$
Ariel G. S\'anchez$^{4,6}$
\\
$^{1}$Argelander Institut f\"ur Astronomie der Universit\"at Bonn, Auf dem H\"ugel 71, 53121 Bonn, Germany\\
$^{2}$Institute of Space Sciences (ICE, CSIC), Campus UAB, Carrer de Can Magrans, s/n, 08193 Barcelona, Spain\\
$^{3}$Institut d’Estudis Espacials de Catalunya (IEEC), 08034 Barcelona, Spain\\
$^{4}$Max-Planck-Institut f\"ur extraterrestrische Physik, Postfach 1312, Giessenbachstr., 85748 Garching, Germany \\
$^{5}$Center for Cosmology and Particle Physics, Department of Physics, New York University, NY 10003, New York, USA\\
$^{6}$Universit\"ats-Sternwarte M\"unchen,  Fakult\"at f\"ur Physik, Ludwig- Maximilians-Universit\"at M\"unchen, Scheinerstrasse 1, 81679 M\"unchen, Germany
}

\date{Accepted XXX. Received YYY; in original form ZZZ}

\pubyear{2022}

\begin{document}
\label{firstpage}
\pagerange{\pageref{firstpage}--\pageref{lastpage}}
\maketitle

\begin{abstract}
  In this paper we present \emu, a Gaussian process emulator of the galaxy power spectrum multipoles in redshift-space. The model predictions are based on one-loop perturbation theory and we consider two alternative descriptions of redshift-space distortions: one that performs a full expansion of the real- to redshift-space mapping, as in recent effective field theory models, and another that preserves the non-perturbative impact of small-scale velocities by means of an effective damping function. The outputs of \emu\ can be obtained at arbitrary redshifts, for arbitrary fiducial background cosmologies, and for a large parameter space that covers the shape parameters $\omega_c$, $\omega_b$, and $n_s$, as well as the evolution parameters $h$, $A_s$, $\Omega_K$, $w_0$, and $w_a$. This flexibility does not impair \emu's accuracy, since we exploit an exact degeneracy between the evolution parameters that allows us to train the emulator on a significantly reduced parameter space. While the predictions are sped up by two orders of magnitude, validation tests reveal an accuracy of $0.1\,\%$ for the monopole and quadrupole ($0.3\,\%$ for the hexadecapole), or alternatively, better than $0.25\,\sigma$ for all three multipoles in comparison to statistical uncertainties expected for the \emph{Euclid} survey with a tenfold increase in volume. We show that these differences translate into shifts in mean posterior values that are at most of the same size, meaning that \emu\ can be used with the same confidence as the exact underlying models. \emu\ is a publicly available \texttt{Python} package that also provides the tree-level bispectrum multipoles and Gaussian covariance matrices.
\end{abstract}

\begin{keywords}
  methods: data analysis; (cosmology:) cosmological parameters; (cosmology:) large-scale structure of Universe 
\end{keywords}

%%%%%%%%%%%%%%%%%%%%%%%%%%%%%%%%%%%%%%%%%%%%%%%%%%%%%%%%%%%%%%%%%%%%%%%%%%%%%%%%%%%%%%%%%%%%%%%%%%%%%%%%%%%%%%%%%%%%%%%%%%%%%%%%%%%%%%%%%%%%%%%%%%%%%%

%%%%%%%%%%%%%%%%% BODY OF PAPER %%%%%%%%%%%%%%%%%%%%%%%%%%%%%%%%%%%%%%%%%%%%%%%%%%%%%%%%%%%%%%%%%%%%%%%%%%%%%%%%%%%%%%%%%%%%%%%%%%%%%%%%%%%%%%%%%%%%%%

\section{Introduction}
\label{sec:introduction}

Models of galaxy clustering statistics inspired by perturbation theory will be one of the main avenues for extracting cosmological information from ongoing and upcoming galaxy surveys, such as the Dark Energy Spectroscopic Instrument \citep[DESI;][]{DESI} and \emph{Euclid} \citep{Euclid}. That is because on sufficiently large scales, where the validity of the perturbative models can be guaranteed, they reach an unparalleled degree of accuracy, making them a very robust and reliable analysis tool. However, even though perturbative methods have been frequently and successfully used in the analysis of past data sets \citep[e.g.][]{SanScoCro1701,BeuSeoSai1704,IvaSimZal2005,dAmGleKok2005,SemSanPez2206}, they cannot be blindly applied to future measurements. The density field associated to each observed population of galaxies will be differently biased with respect to the underlying dark matter field, will have undergone different levels of non-linear evolution, and will be subject to different statistical uncertainties, all of which influence the range of scales on which we can trust the perturbative models. Consequently, each new analysis will need to be accompanied by careful studies of their validity by propagating choices in the modelling, as well as the impact of potential systematics, onto the final results: the cosmological parameters.

Such a task requires running a large number of Monte Carlo Markov Chains (MCMC) in order to determine the posterior parameter distributions, each of which typically takes of the order ${\cal O}(10^6)$ likelihood evaluations for convergence. Although a single evaluation of the perturbation theory models is comparatively fast, even when performing full-shape fits, and can be executed in $\sim 1\,\mathrm{s}$ \citep[e.g.][]{ChuIvaPhi2009,CheVlaCas2103}, this still amounts to considerable computational costs. We intend to greatly facilitate this task by constructing an emulator for various clustering statistics in perturbation theory, in particular the multipoles of the redshift-space galaxy power spectrum, that is capable of drastically reducing the time needed for making model predictions \citep[see][for a similar approach]{DeRCheWhi2204,DonKoyBeu2202}. We thus follow a recent string of works, which introduced emulators as a means of overcoming the much higher computational costs of running simulations, which are needed to extend models of galaxy clustering into the deeply non-linear regime. This enabled efficient modelling of the non-linear matter power spectrum for a set of cosmological parameters \citep{HeiHigWhi0911,KnaStaMar1904,GibCatMoe1912,KnaStaPot2108,AngZenCon2111}, but in combination with a halo occupation distribution model also of the galaxy power spectrum \citep{KwaHeiHab1509,KobNisTak2009}, and the galaxy correlation function in redshift space \citep{ZhaTinBec1903,YuaGarEis2203}. Furthermore, a hybrid approach that makes use of a perturbative expansion of the galaxy bias relation, but obtains all ingredients from simulations was proposed for the galaxy power spectrum  in \citet{ZenAngPel2101,AriAngZen2104,KokDeRChe2107,PelStuAng2208}. While the accuracy of these emulators is limited by the usually small number of training samples, it is straightforward to build large training sets for models purely based on perturbation theory. Our demand is therefore somewhat different: we want a flexible emulator that covers a large space of model parameters, brings large improvements in computation time and yet does not compromise on the principal advantage of perturbation theory, its accuracy.

In order to satisfy that demand it is important to keep the parameter space over which the emulator is build as small as possible. As in \citet{DonKoyBeu2202,AriAngZen2104}, we make use of the fact that all model parameters related to the galaxy bias expansion can be factorised and so their dependency does not need to be emulated. However, what sets our emulator apart from these other works is that it is build around the recently proposed \emph{evolution mapping} approach \citep{SanRuiGon2108}, which further limits the emulation parameter space. In this approach, the cosmological parameters are split into two classes: shape parameters, such as the physical baryon and cold dark matter densities, $\omega_b$, $\omega_c$, and the spectral index $n_s$, and evolution parameters, such as the Hubble parameter $h \equiv H_0/(100\,\mathrm{km}\,\mathrm{s}^{-1}\,\invMpc)$, the scalar amplitude of fluctuations $A_s$, the dark energy equation of state parameters, $w_0$, $w_a$, and the curvature density parameter $\Omega_K$. At any given redshift the evolution parameters can only affect the amplitude of the linear matter power spectrum, provided it is expressed in units of $\Mpc$ (opposed to the conventional $\hMpc$), and thus follow an exact degeneracy \citep{San2012,SanRuiGon2108}. Evolution mapping exploits that degeneracy by mapping models with different evolution parameters from one to the other by simply relabelling the redshifts that correspond to the same value of $\sigma_{12}$, the RMS of density fluctuations in spheres of radius $12\,\Mpc$. We show that this degeneracy extends to the models of perturbation theory considered here, although in redshift-space we additionally need to account for the dependence on the growth rate $f$. Hence, by training our emulator in terms of $\sigma_{12}$ and $f$ in combination with the shape parameters, we can model the effect of the full set of evolution parameters and at arbitrary redshifts. In redshift-space clustering we also need to account for the Alcock-Paczynski (AP) effect \citep{AlcPac7910,BalPeaHea9610}, which requires the assumption of a fiducial background cosmology. We do not include the AP effect in the emulator itself, but in a separate step, and therefore retain more flexibility, which allows us to support any fiducial cosmologies.

This paper is organised as follows: in Sec.~\ref{sec:models} we start by reviewing the perturbative expressions for the galaxy power spectrum multipoles, which our emulator is based on. In particular, we consider two separate models, which differ in their treatment of the real- to redshift-space mapping and show how they can be related. Readers familiar with this material can skip ahead to Sec.~\ref{sec:emulator}, where we describe in detail the evolution mapping approach, as well as further design choices and the computational performance of our emulator. In Sec.~\ref{sec:validation} we conduct a series of stringent validation tests of our emulator and propagate any emulation uncertainties onto the posterior distributions of the model parameters, demonstrating no relevant loss in accuracy. We conclude in Sec.~\ref{sec:conclusions} with a comparison to related works and an overview of the capabilities of our \texttt{Python} package \emu.

%%%%%%%%%%%%%%%%%%%%%%%%%%%%%%%%%%%%%%%%%%%%%%%%%%%%%%%%%%%%%%%%%%%%%%%%%%%%%%%%%%%%%%%%%%%%%%%%%%%%%%%%%%%%%%%%%%%%%%%%%%%%%%%%%%%%%%%%%%%%%%%%%%%%%%
\section{Overview of perturbation theory models}
\label{sec:models}

In this work, we consider the emulation of the redshift-space power spectrum multipoles of biased tracers for two different perturbative models. These models differ solely in their treatment of redshift-space distortions with one performing a full expansion of the real-to-redshift space mapping, while the other partly retains the non-perturbative nature of this mapping. Examples of the first category of models are the recent effective field theory (EFT) models of \cite{IvaSimZal2005} and \cite{dAmGleKok2005}, which introduce a series of counterterms that are meant to capture the effect from small-scale, virialised velocities of galaxies on the power spectrum. Models, such as TNS \citep{TarNisSai1009} or the one proposed by \cite{SanScoCro1701}, fall into the second category and account for the virialised velocity impact via an effective damping function that represents the velocity difference generating function (VDG). For the remainder of this paper we will refer to these two models as EFT and VDG models, respectively.

As an overview, the galaxy power spectrum in the EFT model is generally decomposed into four different terms,
\begin{equation}
  \label{eq:PEFT}
  P_{gg,\rm EFT}(\bk) = P_{gg,\rm SPT}^{\rm tree}(\bk) + P_{gg,\rm SPT}^{\rm 1-loop}(\bk) + P_{gg}^{\rm stoch}(\bk) + P_{gg}^{\rm ctr}(\bk)\,,
\end{equation}
where the first two refer to the leading and next-to-leading (tree-level and one-loop, respectively) contributions from standard perturbation theory (SPT), which result from an expansion of the real to redshift-space mapping, the non-linear evolution of the matter and velocity fields, and galaxy bias (see Sec.~\ref{sec:RSD_power} and \ref{sec:bias}). The third term accounts for stochastic effects due to the impact from highly non-linear scales on the galaxy bias relation, while the last term represents the series of counterterms (see Sec.~\ref{sec:HD+stoch}) that are meant to absorb any sensitivities of the one-loop term to modes beyond the reach of perturbation theory and, as stated before, the effect from virialised velocities. The above expression is furthermore supplemented by a procedure (infrared resummation) that takes into account the damping of baryon acoustic oscillations (BAO) from large-scale relative displacements, the details of which will be presented in Sec.~\ref{sec:IR-resummation}. The galaxy power spectrum in the VDG model involves all the same ingredients, and we show that it can be related to the EFT model as follows:
\begin{equation}
  \label{eq:PVDG}
  P_{gg, \rm VDG}(\bk) = W_{\infty}(\bk)\,\left[P_{gg, \rm EFT}(\bk) - \Delta P(\bk)\right]\,,
\end{equation}
where $W_{\infty}$ denotes the effective damping function and $\Delta P(\bk)$ is a correction term that stems from the fact that not the full real to redshift-space mapping is expanded perturbatively and will be computed in Sec.~\ref{sec:RSD_power}.

Throughout this section and for the remainder of this paper we assume the plane-parallel approximation for redshift-space distortions\footnote{This means that all pairs of galaxies have the same line-of-sight, which we take to be the $\hat{z}$-direction.} and work under the Fourier transform convention
\begin{equation}
  \label{eq:FourierConvention}
  \delta(\bx) = \int_{\bk} \mathrm{e}^{-i \bk \cdot \bx}\,\delta(\bk)\,.
\end{equation}
All $\bk$-space integrals are written using the short-hand notation $\int_{\bk_1,\ldots\,,\bk_n} \equiv \int \mathrm{d}^3k_1/(2\pi)^3 \cdots \mathrm{d}^3k_n/(2\pi)^3 $, and in configuration space we analogously use $\int_{\bx_1,\ldots\,,\bx_n} \equiv \int \mathrm{d}^3x_1 \cdots \mathrm{d}^3x_n$.

\subsection{Redshift-space power spectra in EFT and VDG models}
\label{sec:RSD_power}

The mapping from real-space position $\bx$ to redshift-space position $\bs$ in the plane-parallel approximation is given by: 
\beq
\bs=\bx - f \ u_z(\bx) {\hat z}, 
\label{RSDmap}
\eeq 
where $u_z$ is the line-of-sight peculiar velocity normalised such that $\theta=\nabla\cdot \bu=\delta$ in linear theory (i.e. $\bv\equiv -f {\cal H} \bu$), $f$ is the growth rate and ${\cal  H}$ the conformal Hubble rate. This yields for the galaxy density perturbations in redshift-space~\citep{ScoCouFri9906},
\beq
  \delta_s(\bk,z) = \int_{\bx} \mathrm{e}^{i \bk \cdot \bx} {\mathrm e}^{-i f k_z u_z(\bx)} D_s(\bx)
 \label{deltaS}
 \eeq 
where we assumed that the Jacobian of the map in Eq.~(\ref{RSDmap}) can be approximated $J=|1-f \nabla_z u_z| \simeq 1-f \nabla_z u_z$  and $D_s(\bx) \equiv \delta_g(\bx) + f \nabla_z u_z(\bx)$ is a combination of the galaxy density contrast $\delta_g$ and the gradient of $u_z$ along the line-of-sight. Expanding $\delta$ and $\bu$ in perturbation theory, yields the perturbation theory kernels in redshift-space $Z_n$~\citep{ScoCouFri9906}
\begin{equation}
  \label{eq:Zn_expansion}
  \begin{split}
  \delta_s(\bk,z) = \sum_{n=1} D^n(z) \int_{\bk_1,\ldots\,,\bk_n} \delta_D(\bk - \bk_{1 \ldots n})\,Z_n(\bk_1,\ldots\,,\bk_n) \\ \times\,\delta_L(\bk_1) \cdots \delta_L(\bk_n)\,,
  \end{split}
\end{equation}
which not only contain information about the redshift-space distortions, but also about galaxy bias, which we describe next in Sec.~\ref{sec:bias} (see Appendix~\ref{sec:loop_integrals_expressions} for explicit expressions). This leads to the standard perturbation theory expressions for the power spectrum in redshift-space at leading and next-to-leading (one-loop) order,
\begin{align}
  P_{gg,\rm SPT}^{\rm tree}(\bk) &= Z_1(\bk)^2\,P_L(k) \label{eq:Ptree}\\
  P_{gg,\rm SPT}^{\rm 1-loop}(\bk) &= 2 \int_{\bq}  \left[Z_2(\bk-\bq,\bq)\right]^2 P_L(|\bk-\bq|) P_L(q) \nonumber \\ & \hspace{0.85em}+6\,P_L(k) \int_{\bq} Z_3(\bk,\bq,-\bq) P_L(q)   \label{eq:P1Lspt}
\end{align}
where $\bk$ is described by its magnitude $k$ and its cosine $\mu$ with respect to the fixed line-of-sight in the plane-parallel approximation, $\mu \equiv \hat{\bk}\cdot\hat{z}$, and $P_L$ is the linear power spectrum in which we have absorbed the redshift dependence.

In order to relate the power spectrum in the VDG model to the SPT computation we start from the exact expression of the redshift-space galaxy power spectrum that can be derived from the redshift-space map in Eq.~(\ref{RSDmap}), written in terms of the pairwise velocity generating function ${\cal M}(\lambda,\boldsymbol{r})$ \citep[see][for a derivation]{Sco0410}:
\begin{equation}
  \label{eq:Pzs}
  P_{gg}(\bk) = \int_{\boldsymbol{r}} \mathrm{e}^{i \boldsymbol{k} \cdot \boldsymbol{r}}\,\Big\{\big[1 + \xi_{gg}(r)\big]\,{\cal M}(\lambda,\boldsymbol{r}) - 1\Big\}\,,
\end{equation}
where $\xi_{gg}(r)$ is the real-space galaxy two-point correlation function at separation $r$ and $\lambda \equiv - i\,f k \mu$. As shown in \cite{Sco0410,TarNisSai1009}, the pairwise velocity generating function can be decomposed into the following building blocks of connected correlators:
\begin{equation}
  \label{eq:M}
  \begin{split}
  \left[1 + \xi_{gg}(r)\right]\,{\cal M}(\lambda,\boldsymbol{r}) =W(\lambda,\boldsymbol{r}) \left[\left<\mathrm{e}^{\lambda\,\Delta u_z}\,D_s(\boldsymbol{x})\,D_s(\boldsymbol{x}')\right>_c \right. \\ \left. + \left<\mathrm{e}^{\lambda\,\Delta u_z}\,D_s(\boldsymbol{x})\right>_c\left<\mathrm{e}^{\lambda\,\Delta u_z}\,D_s(\boldsymbol{x}')\right>_c\right]\,,
  \end{split}
\end{equation}
where $\Delta u_z$ is the difference in line-of-sight peculiar velocities at positions $\boldsymbol{x}$ and $\boldsymbol{x}'=\boldsymbol{x} + \boldsymbol{r}$. The prefactor $W(\lambda,\boldsymbol{r}) \equiv \exp{\left<\mathrm{e}^{\lambda\,\Delta u_z}\right>_c}$ is the velocity difference generating function. The velocity difference at separation $r$ is most sensitive to small-scale modes, since large-scale modes with $k r\lesssim1$ contribute the same velocity to both $\boldsymbol{x}$ and $\boldsymbol{x}'$. Therefore the VDG model attempts to keep this term non-perturbative and instead replaces it by an effective damping function, calculated in the large-scale limit ($r\to\infty$), given by~\citep{SanScoCro1701}
\begin{equation}
  \label{eq:Winfty}
  W_{\infty}(\lambda) = \frac{1}{\sqrt{1- \lambda^2\,a_{\rm vir}^2}}\,\exp{\left(\frac{\lambda^2\,\sigma_v^2}{1 - \lambda^2\,a_{\rm vir}^2}\right)}\,,
\end{equation}
where $a_{\rm vir}$ is a free parameter that contributes to small-scale velocity dispersion and controls the kurtosis of the VDG, i.e. the limit $a_{\rm vir}=0$ corresponds to Gaussian velocity differences. The fact that $a_{\rm vir}\neq 0$ reflects the well-known result that even in the large-scale limit the probability distribution function  of pairwise velocities is non-Gaussian with significant exponential tails~\citep{She9604,JusFisSza9809,Sco0410,CueLiEgg2010}. 

In the VDG model, the remaining terms in square brackets in Eq.~(\ref{eq:M}) are treated perturbatively, i.e. expanding the exponentials to give an expression consistent to one-loop order. As in the EFT model this expansion includes galaxy bias and stochastic terms, as well as counterterms from small-scale modes, while all resulting expressions are evaluated using BAO-damped linear power spectra as described in Section~\ref{sec:IR-resummation}. Since Eqs.~(\ref{eq:Pzs}-\ref{eq:M}) are of course compatible with perturbation theory, the relation between the EFT and VDG power spectra is found by computing only the additional terms $\Delta P(\bk)$ that result from the expansion of the velocity difference generating function. Up to one-loop order it is sufficient to expand the exponential as $\exp{\left<\mathrm{e}^{\lambda\,\Delta u_z}\right>_c} \approx 1 + \lambda^2/2 \left<\Delta u_z^2\right>$ , which leads to one extra term given by
\begin{align}
  \label{eq:DeltaP}
  \Delta P(k,\mu) &= \int_{\boldsymbol{r}} \mathrm{e}^{i \bk \cdot \boldsymbol{r}}\,\left[\frac{\lambda^2}{2} \left<\Delta u_z^2\right> \left<D_s(\bx)\,D_s(\bx')\right>\right] \nonumber \\
  &= \lambda^2\,\sigma_v^2\,P_{D_s D_s}(k,\mu) - \lambda^2 \int_{\qv} \frac{q_z^2}{q^4} P_{\theta \theta}(q)\,P_{D_s D_s}(\bk-\qv)\,,
\end{align}
where $P_{D_s D_s}$ and $P_{\theta \theta}$ are the power spectra of the $D_s(\bx)$ field and the velocity divergence, both of which are evaluated at leading order in Eq.~(\ref{eq:DeltaP}), which implies $P_{D_s D_s}(\bk) = P_{gg,\rm SPT}^{\rm tree}(\bk)$. The linear velocity dispersion, $\sigma_v$, appearing in the first term on the second line of Eq.~(\ref{eq:DeltaP}) is given by:
\begin{equation}
  \label{eq:sigmav}
  \sigma_v^2 = \frac{1}{3} \int_{\bk} \frac{P_{\theta\theta}(k)}{k^2} = \frac{1}{3} \int_{\bk} \frac{P_{L}(k)}{k^2} \,.
\end{equation}
Finally, since $W_{\infty}$ does not depend on scale $\boldsymbol{r}$, we can pull it out of the integral in Eq.~(\ref{eq:Pzs}), from which Eq.~(\ref{eq:PVDG}) follows.

We note that the VDG model as presented here differs from the TNS model~\citep{TarNisSai1009} in the effective damping function $W_{\infty}$ and in that the square bracket in Eq.~(\ref{eq:M}) includes all contributions from quadratic bias, both local and nonlocal~\citep{SanScoCro1701}.

\subsection{Galaxy bias expansion and stochasticity}
\label{sec:bias}

In perturbation theory we predict the clustering of biased tracers, such as dark matter halos or galaxies, by first relating their over-densities to a series of properties of the underlying dark matter field. This is known as the galaxy bias expansion (see \citet{DesJeoSch1802} for a comprehensive review), where each property (in the following also denoted as \emph{operator}) captures an effect of the large-scale environment on the formation and evolution of galaxies, and the strength of that effect is encoded in the associated galaxy bias parameters. While the values of the bias parameters cannot be computed from first principles and also depend on the selection of the observed population of tracers, the various operators to be taken into account at each order of perturbation theory can be derived based on symmetry considerations \citep{McDRoy0908,ChaSco1204,AssBauGre1408,Sen1511,MirSchZal1507,DesJeoSch1802,EggScoSmi1906}.

Choosing the set of operators proposed in \citet{EggScoSmi1906}, and showing only quantities that are relevant for the computation of the power spectrum at next-to-leading order, results in the following expansion:
\begin{equation}
  \label{eq:deltag_expansion}
  \begin{split}
  \delta_g =\;&b_1\,\delta + \frac{b_2}{2}\,\delta^2 + \gamma_2\,{\cal G}_2(\Phi_v) + \gamma_{21}\,{\cal G}_{21}(\varphi_2,\varphi_1) \\ &+ b_{\nabla^2}\nabla^2\delta + \epsilon_g + \ldots\,,
  \end{split}
\end{equation}
with each term being evaluated at position $\bx$ and redshift $z$. The two Galileon operators ${\cal G}_2$ and ${\cal G}_{21}$, which quantify the impact of large-scale tides at second and third order, are defined as (sums over repeated indices are implied)
\begin{align}
  \label{eq:Galileons}
  {\cal G}_2(\Phi_v) &\equiv \nabla_i \nabla_j \Phi_v\,\nabla_i \nabla_j \Phi_v - \left(\nabla^2 \Phi_v\right)^2\,, \\
  {\cal G}_{21}(\varphi_2,\varphi_1) &\equiv \nabla_i \nabla_j \varphi_2\,\nabla_i \nabla_j \varphi_1 - \nabla^2\varphi_2\,\nabla^2\varphi_1\,,
\end{align}
where $\Phi_v = \nabla^{-2}\theta$ is the velocity potential and $\varphi_i$ are the potentials of Lagrangian perturbation theory, satisfying:
\begin{align}
  \nabla^2\varphi_1 = -\delta\,, \quad \nabla^2\varphi_2 = -{\cal G}_2(\varphi_1)\,.
\end{align}
Transformed into Fourier space, these relations give rise to the galaxy bias kernel functions given in Eqs.~(\ref{eq:K2}) and (\ref{eq:K3}).

There are a number of other, equivalent galaxy bias bases in the literature with a different choice of operators, and their corresponding bias parameters can be mapped into those described above via simple linear transformations. For instance, the basis used by \cite{AssBauGre1408,IvaSimZal2005} contains two different operators and their bias parameters are related to the ones here as
\begin{equation}
  \label{eq:bias_relation_Iva}
  \gamma_2 = b_{{\cal G}_2}\,, \quad \gamma_{21} = -\frac{4}{7}\left(b_{{\cal G}_2} + b_{\Gamma_3}\right)\,,
\end{equation}
whereas the four bias parameters defined in \cite{dAmGleKok2005} (denoted here with a tilde) convert into ours as follows:
\begin{equation}
  \label{eq:bias_relation_Ami}
  \begin{split}
  &b_1 = \tilde{b}_1\,, \quad b_2 = 2\,\left(-\tilde{b}_1 + \tilde{b}_2 + \tilde{b}_4\right)\,, \quad \gamma_2 = -\frac{2}{7} \left(\tilde{b}_1 - \tilde{b}_2\right)\,, \\
  &\gamma_{21} = -\frac{2}{147} \left(11 \tilde{b}_1 - 18 \tilde{b}_2 + 7\tilde{b}_3 \right)\,.
  \end{split}
\end{equation}
While the default in \emu\ is the expansion based on Eq.~(\ref{eq:deltag_expansion}), it is possible to specify bias parameters corresponding to either of the other two bases above.

The last two terms in Eq.~(\ref{eq:deltag_expansion}) represent the relevant contributions from higher-derivative operators and stochasticity, respectively. The former arise when taking into account that gravitational collapse occurs within patches of finite size $R$, meaning that on distance scales of comparable magnitude, the galaxy over-density at position $\bx$ can no longer be assumed to depend on the matter properties at the \emph{same} position alone. Expanding this spatial non-locality perturbatively leads to terms scaling as $\sim R^2\,\nabla^2 {\cal O}$ with ${\cal O}=\{\delta\,,\delta^2\,,{\cal G}_2\,,\ldots\}$ \citep{Des0811,McDRoy0908,DesCroSco1011,DesJeoSch1802} and assuming that the scale $R$ is of similar order as the non-linearity scale of the matter field, the term $\nabla^2\delta$ is the only one relevant for the power spectrum at next-to-leading order. Its bias parameter does not appear in the kernels in Eqs.~(\ref{eq:K2}) and (\ref{eq:K3}), because in Sec.~\ref{sec:HD+stoch} we will absorb it into the definition of one of the counterterms that carries the same scale dependence.

Finally, the stochastic field $\epsilon_g$ is meant to summarise all highly non-linear effects on galaxy formation that are not captured by the perturbative expansion. The corresponding wavemodes are uncorrelated with those accessible in perturbation theory and thus all of the large-scale fields appearing in Eq.~(\ref{eq:deltag_expansion}), meaning that on large scales their contribution appears stochastic \citep{DekLah9907,TarSod9909,Mat9911}. They therefore enter the galaxy power spectrum as an independent term, the stochastic power spectrum $\left<\epsilon_g(\bk_1)\,\epsilon_g(\bk_2)\right> = (2\pi)^3\,P_{\epsilon_g\,\epsilon_g}(k_1) \delta_D(\bk_{12})$, which can be expanded in the low-$k$ regime as \citep{DesJeoSch1802,EggScoSmi2106}:
\begin{equation}
  \label{eq:Pshot}
  P_{\epsilon_g\epsilon_g}(k) = \frac{1}{\bar{n}}\left(N^P_{0} + N^P_{2,0}\,k^2 + \ldots\right)\,,
\end{equation}
where $\bar{n}$ denotes the mean number density of the tracer population and we have defined the two stochastic bias parameters, $N^P_{0}$ and $N^P_{2,0}$. Through the redshift-space mapping the galaxy power spectrum is also sensitive to a stochastic component in the gradient of the line-of-sight velocity field, $\epsilon_{\nabla_z u_z}$, which leads to one additional term \citep{PerSenJen1610,DesJeoSch1812} that was shown to account for the small-scale velocity dispersion \citep{DesJeoSch1812}. We parametrise this term as follows:
\begin{equation}
  \label{eq:Pshot_vel}
  P_{\epsilon_g\epsilon_{\nabla_z u_z}}(k,\mu) = \frac{N^P_{2,2}}{\bar{n}} {\cal L}_2(\mu)\,k^2 + \ldots\,,
\end{equation}
where ${\cal L}_2(\mu)$ denotes the second Legendre polynomial\footnote{Using the Legendre polynomial in the definition of $N^P_{2,2}$ is an arbitrary choice, which simply guarantees that this term can only contribute to the quadrupole of the power spectrum in the absence of Alcock-Paczynski distortions.}, such that the total stochastic contribution to the galaxy power spectrum is given by
\begin{equation}
  \label{eq:Pstoch}
  P_{gg}^{\rm stoch}(k,\mu) = P_{\epsilon_g\epsilon_g}(k) + P_{\epsilon_g\epsilon_{\nabla_z u_z}}(k,\mu)\,.
\end{equation}
Since in the VDG model the impact from small-scale velocities is encoded in the effective damping function $W_{\infty}$ and not expanded perturbatively, it is likely that the second term in Eq.~(\ref{eq:Pstoch}) is of subordinate importance in that case.

\subsection{Definition of counterterms}
\label{sec:HD+stoch}

The loop integrals in Eq.~(\ref{eq:P1Lspt}) or Eq.~(\ref{eq:DeltaP}) are performed over the entire range of scales, even those where perturbation can no longer be applied. In order to construct a consistent theory, one needs to introduce a series of \emph{counterterms} with adjustable amplitudes (i.e., a free parameter per counterterm), which are able to absorb any sensitivity to non-linear modes that might arise in the large-scale limit. It was shown e.g. in \cite{SenZal1409,DesJeoSch1812} that the required leading counterterms for the galaxy power spectrum in redshift-space scale as $\sim \mu^{2n}\,k^2\,P_L(k)$ with $n=0\,,1\,,2$, assuming them for simplicity to be local in time. The first of these scales exactly like the higher-derivative bias contribution $\nabla^2\delta$ (see Sec.~\ref{sec:bias}) and so we can absorb its coefficient, $b_{\nabla^2\delta}$, into the corresponding counterterm parameter. The same counterterm also absorbs the leading effect from a breakdown of the perfect fluid approximation for the matter field \citep{PueSco0908,BauNicSen1207,CarHerSen1209}. Moreover, \citet{DesJeoSch1812} demonstrated that the other two counterterms with $n=1$ and $n=2$ can account for the relevant velocity bias effects, which we have neglected in our description so far.

In total, we therefore introduce another three free parameters, $c_0$, $c_2$ and $c_4$, and define the contribution from the leading order counterterms to the galaxy power spectrum as follows:
\begin{equation}
  \label{eq:PctrLO}
  P_{gg}^{\rm ctr, LO}(k,\mu) = -2 \sum_{n=0}^{2} c_{2n}\,{\cal L}_{2n}(\mu)\,k^2\,P_L(k)\,,
\end{equation}
where instead of the scaling with $\mu^{2n}$ we used Legendre polynomials of order $2n$. This simply corresponds to a linear transformation of the counterterm parameters, and means that each of them predominantly contributes to only a single power spectrum multipole.

The perturbative expansion of the velocity difference generating function in case of the EFT model can also be regarded as a higher-derivative expansion, scaling roughly as $\sim (\mu\,k\,\sigma_v)^{2n}\,P_{gg,\rm SPT}^{\rm tree}(\bk)$, with Eq.~(\ref{eq:DeltaP}) representing the term relevant at the one-loop level ($n=1$). However, the length scale associated with this expansion is $\sigma_v^{-1}$, which could potentially be significantly different from the non-linearity scale of the matter field. In particular, if $\sigma_v^{-1}$ is smaller than the non-linearity scale, this would imply that terms relevant at higher than one-loop should be retained in that expansion as well. For that reason, \citet{IvaSimZal2005} include the next-order term in the model for the galaxy power spectrum with a free amplitude that is meant to marginalise over such potential contributions:
\begin{equation}
  \label{eq:PctrNLO}
  P_{gg}^{\rm ctr,NLO}(k,\mu) = c_{\rm nlo}\,(\mu\,k\,f)^4\,P_{gg,\rm SPT}^{\rm tree}(\bk)\,,
\end{equation}
so that the full counterterm power spectrum reads:
\begin{equation}
  \label{eq:Pctr}
  P_{gg}^{\rm ctr}(k,\mu) = P_{gg}^{\rm ctr,LO}(k,\mu) + P_{gg}^{\rm ctr,NLO}(k,\mu)\,.
\end{equation}
Since the NLO term is purely coming from the expansion of the velocity difference generating function, we do not include it in case of the VDG model.

\subsection{Infrared resummation}
\label{sec:IR-resummation}

The model presented in the previous sections provides a good description of the broadband of the anisotropic galaxy power spectrum on mildly non-linear scales, but exhibits a non negligible difference in terms of the amplitude of the BAO wiggles \citep[e.g.][]{BalMirSim1508}. The latter are most sensitive to large-scale bulk flows, which are responsible for the smearing of the BAO signal via the large-scale relative displacement field. 

Following a perturbative approach, corrections to the matter power spectrum can be resummed at each wavemode $k$ by taking into account the effect of fluctuations on larger scales, i.e. at $q<k$. At leading order, the net effect on the matter power spectrum is that of a damping factor, which only acts on the BAO wiggles. From a practical point of view, it is standard practice to split the linear matter power spectrum into a smooth and wiggly component \citep{BalMirSim1508,BlaGarIva1607}, such that
\begin{equation}
    \Plin(k)=\Pnw(k)+\Pw(k).
\end{equation}
The wiggle/no-wiggle split is carried out making use of the same implementation suggested in \cite{VlaSelChu1603,OsaNisBer1903}. This reads
\begin{equation}
    \Pnw(k)=\Peh(k)\,\mathcal{F}\left[\frac{\Plin(k)}{\Peh(k)}\right],
\end{equation}
and consists in a rescaling of the original formula for the featureless matter power spectrum $\Peh(k)$ first presented in \cite{EisHu9804}, and adjusted to match the broadband amplitude of the linear matter power spectrum. We choose the band filter $\mathcal{F}$ to be Gaussian, with its functional form defined by
\begin{equation}
    \begin{split}
        \mathcal{F}\left[f(k)\right] = \;& \frac{1}{\sqrt(2\pi)\log_{10}(\lambda)}\int {\rm{d}}\left(\log_{10}(q)\right) f(q)\, \times \\ &\times\exp\left[-\frac{\left(\log_{10}(k)-\log_{10}(q)\right)^2}{2\left(\log_{10}(\lambda)\right)^2}\right],
    \end{split}
\end{equation}
where $\log_{10}\left(\lambda/(h^{-1}\mathrm{Mpc})\right)=0.25$.

At leading order, the infrared-resummed matter power spectrum can then be written as the sum of the smooth component and the damped wiggly one, such that \citep{BalMirSim1508,BlaGarIva1607}
\begin{equation}
    P_{\rm{mm}}^{\,\rm{IR-LO}}(k) = \Pnw(k)+e^{-k^2\Sigma^2}\Pw(k),
\end{equation}
where
\begin{equation}
    \Sigma^2=\frac{1}{6\pi^2}\int_0^{\ks} \Pnw(q)\left[1-j_0\left(\frac{q}{\kosc}\right)+2j_2\left(\frac{q}{\kosc}\right)\right]{\rm{d}}q
\end{equation}
is the two point function of the relative displacement field evaluated at the BAO scale. Here, $j_n$ is the $n$-th order spherical Bessel function, $\kosc=1/\ell_{\rm{osc}}$ is the wavemode corresponding to the reference BAO scale $\ell_{\rm{osc}}=110\,h^{-1}{\rm{Mpc}}$, and $\ks$ is the ultraviolet integration limit. Even though in principle the resummation of modes should be carried out for all $q<k$, it has become standard practice to fix this upper limit to an intermediate scale. In our implementation we adopt\footnote{Note that the value for $\ks$ is defined in units of $\invMpc$ opposed to $\hinvMpc$ since \emu\ internally works in $\Mpc$ units, see Sec.~\ref{sec:emulator}.} $\ks=0.14\,\mathrm{Mpc}^{-1}$.

At next-to-leading order, the IR-resummed matter power spectrum receives additional contributions, along with the standard one-loop corrections sourced by higher powers of the density field. The final expression then reads \citep{BalMirSim1508,BlaGarIva1607}
\begin{equation}
    \begin{split}
        P_{\rm{mm}}^{\,\rm{IR-NLO}}(k) = \; & \Pnw(k) + \left(1+k^2\Sigma^2\right)e^{-k^2\Sigma^2}\Pw(k) \, +\\
        & + P^{\rm{1-loop}}\left[P_{\rm{mm}}^{\,\rm{IR-LO}}\right](k),
    \end{split}
    \label{eq:pmm_IRNLO}
\end{equation}
where the square brackets in the second line of Eq. (\ref{eq:pmm_IRNLO}) indicate that the one-loop integrals are evaluated using the leading order IR-resummed power spectrum instead of the linear power spectrum.

When considering the redshift-space galaxy power spectrum, the most notable difference is that the damping factor now also carries a dependence on the line-of-sight, such that, at leading order, we can write \citep{IvaSib1807}
\begin{equation}
    P_{\rm{gg}}^{\,s,\rm{IR-LO}}(k,\mu)=\left(b_1+f\mu^2\right)^2\left[\Pnw(k)+e^{-k^2\Sigma_{\rm{tot}}^{\,2}(\mu)}\Pw(k)\right].
\end{equation}
The angular dependence of the new damping factor can be expressed as
\begin{equation}
  \label{eq:IR.sigma_tot}
    \Sigma_{\rm{tot}}^{\,2}(\mu) = \left[1+f\mu^2(2+f)\right]\Sigma^2 + f^2\mu^2(\mu^2-1){\rm{d}}\Sigma^2,
\end{equation}
where
\begin{equation}
    {\rm{d}}\Sigma^2=\frac{1}{2\pi^2}\int_0^{\ks}\Pnw(q)\,j_2\left(\frac{q}{\kosc}\right){\rm{d}}q.
\end{equation}
At next-to-leading order, we can then write \citep{IvaSib1807}:
\begin{equation}
    \begin{split}
        &P_{\rm{gg}}^{\,s,\rm{IR-NLO}}(k,\mu) = \\
        &\left(b_1+f\mu^2\right)^2\left[\Pnw(k)+\left(1+k^2\Sigma_{\rm{tot}}^{\,2}(\mu)\right)e^{-k^2\Sigma_{\rm{tot}}^{\,2}(\mu)}\Pw(k)\right] + \\
        &\,+ P_{\rm{gg}}^{\,s,{\rm{1-loop}}}\left[\Pnw\right](k) + \\ &\,+ e^{-k^2\Sigma_{\rm{tot}}^{\,2}(\mu)}\left(P_{\rm{gg}}^{\,s,{\rm{1-loop}}}\left[\Pnw+\Pw\right](k)-P_{\rm{gg}}^{\,s,{\rm{1-loop}}}\left[\Pnw\right](k)\right),
    \end{split}
\end{equation}
where, once again, square brackets indicate that we are evaluating the one-loop terms by either plugging in the total linear matter power spectrum ($\Pnw+\Pw$) or just the smooth component ($\Pnw$).

%%%%%%%%%%%%%%%%%%%%%%%%%%%%%%%%%%%%%%%%%%%%%%%%%%%%%%%%%%%%%%%%%%%%%%%%%%%%%%%%%%%%%%%%%%%%%%%%%%%%%%%%%%%%%%%%%%%%%%%%%%%%%%%%%%%%%%%%%%%%%%%%%%%%%%
\section{Emulator design}
\label{sec:emulator}

Our emulator is designed around a number of key ideas that minimise the emulation parameter space, while simultaneously keeping the emulator highly flexible, as well as applicable to arbitrary fiducial background cosmologies and a continuous range of redshifts. This is achieved by 1) employing the \emph{evolution mapping} approach of \citet{SanRuiGon2108}, 2) an exact treatment of Alcock-Paczynski distortions, and 3) a separate emulation of all contributions that are proportional to a unique combination of galaxy bias parameters. In the following we describe these ideas and further design choices in detail.

\subsection{Evolution mapping}
\label{sec:evolution_mapping}

The evolution mapping approach classifies the dependence on cosmological parameters into two groups: shape parameters, $\boldsymbol{\Theta}_{\rm s}$, which determine the shape of the linear power spectrum $P_L(k|z)$, and evolution parameters, $\boldsymbol{\Theta}_{\rm e}$, which at any given redshift only affect its amplitude. Examples of the former include the physical densities of radiation, baryons, cold dark matter and neutrinos, as well as the scalar spectral index,
\begin{equation}
  \label{eq:emu.shape_params}
  \boldsymbol{\Theta}_{\rm s} = \big(\omega_{\gamma},\,\omega_{b},\,\omega_{c},\,\omega_{\nu},\,n_s,\,\ldots\big)\,,
\end{equation}
while the latter class contains the scalar amplitude of the primordial power spectrum, the curvature and dark energy density parameters, as well as any parameters that describe the time evolution of the dark energy equation of state,
\begin{equation}
  \label{eq:emu.evo_params}
  \boldsymbol{\Theta}_{\rm e} = \big(A_s,\,\omega_K,\,\omega_{\rm DE},\,w_{\rm DE}(z),\,\ldots\big)\,.
\end{equation}
Crucially, it was shown in \citet{San2012, SanRuiGon2108} that when the linear power spectrum is expressed in units of $\Mpc$ (opposed to the traditional $\hMpc$), all evolution parameters follow an exact degeneracy. This means that their joint effect can be summarised by the value of a single parameter, with a convenient choice being $\sigma_{12}$, the RMS of matter fluctuations in spheres of radius $R = 12\,\Mpc$, as suggested by \citet{SanRuiGon2108}. The evolution mapping relation can then be expressed as
\begin{equation}
  \label{eq:emu.evo_mapping}
  P_L\big(k|z,\boldsymbol{\Theta}_{\rm s},\boldsymbol{\Theta}_{\rm e}\big) = P_L\big(k|\boldsymbol{\Theta}_{\rm s},\sigma_{12}(z,\boldsymbol{\Theta}_{\rm s},\boldsymbol{\Theta}_{\rm e})\big)\,,
\end{equation}
which states that at fixed values of $\boldsymbol{\Theta}_{\rm s}$ different choices of $\boldsymbol{\Theta}_{\rm e}$ can be transformed into each other by relabelling the redshifts that correspond to the same value of $\sigma_{12}$.

We can extend this concept to the perturbation theory models for biased tracers in redshift space that were discussed in Sec.~\ref{sec:models}. These models can be split into individual terms $P_{{\cal B}}(\boldsymbol{k}|z)$ that are proportional to different combinations of galaxy bias parameters ${\cal B}$. Each of these terms is a functional of the linear power spectrum and depends otherwise only on the growth rate of structures $f(z)$.\footnote{The growth rate cannot be fully factorised because it also affects the damping term in the IR resummation procedure (see Sec.~\ref{sec:IR-resummation}), in particular Eq.~(\ref{eq:IR.sigma_tot}).} For that reason and using Eq.~(\ref{eq:emu.evo_mapping}) we can write
\begin{equation}
  \label{eq:emu.evo_nonlinear}
  \begin{split}
    P_{{\cal B}}(\boldsymbol{k}|z,\boldsymbol{\Theta}_{\rm s},\boldsymbol{\Theta}_{\rm e}) = P_{{\cal B}}\big\{&\boldsymbol{k}|f(z,\boldsymbol{\Theta}_{\rm s},\boldsymbol{\Theta}_{\rm e}), \big. \\ & \big.P_L\big(q|\boldsymbol{\Theta}_{\rm s},\sigma_{12}(z,\boldsymbol{\Theta}_{\rm s},\boldsymbol{\Theta}_{\rm e})\big)\big\}\,,
  \end{split}
\end{equation}
implying that the cosmology and redshift dependence of the perturbation theory models can be fully captured in terms of the shape parameters, the growth rate and $\sigma_{12}$. We will exploit this observation to make our emulator applicable to an arbitrary number of evolution parameters and for a continuous range of redshifts, thus giving it a high degree of flexibility.

\vspace{-1em}
\subsection{Application of Alcock-Paczynski distortions and VDG damping function}
\label{sec:emu_AP+vir}

The analysis of galaxy clustering data sets requires the assumption of a fiducial background cosmology, which is used to convert the measured redshifts and angular positions of the galaxies into comoving distance scales. Any differences between the true and fiducial cosmological parameters lead in general to anisotropic distortions of the power spectrum, commonly denoted as Alcock-Paczynski (AP) distortions. They are described by two parameters that quantify the rescaling of the distance scales parallel and perpendicular to the line-of-sight and are given by the ratios of the Hubble parameter and the comoving transverse distance in the true and fiducial cosmologies at the mean redshift of the galaxy sample\footnote{We note that depending on the chosen unit system ($\Mpc$ vs. $\hMpc$), $H$ and $D_M$ need to be given either in units of $\mathrm{km}\,\mathrm{s}^{-1}\Mpc^{-1}$ and $\Mpc$, or $\mathrm{km}\,\mathrm{s}^{-1}(\hMpc)^{-1}$ and $\hMpc$, respectively.}:
\begin{align}
  q_{\parallel} &\equiv \frac{H(z_m|\boldsymbol{\Theta}_{\rm s,fid},\boldsymbol{\Theta}_{\rm e,fid})}{H(z_m|\boldsymbol{\Theta}_{\rm s},\boldsymbol{\Theta}_{\rm e})} \\
  q_{\perp} &\equiv \frac{D_M(z_m|\boldsymbol{\Theta}_{\rm s},\boldsymbol{\Theta}_{\rm e})}{D_M(z_m|\boldsymbol{\Theta}_{\rm s,fid},\boldsymbol{\Theta}_{\rm e,fid})}\,.
\end{align}
They are determined by a combination of the shape and evolution parameters, such that following the approach of Sec.~\ref{sec:evolution_mapping} it is not possible to include the AP effect implicitly (which is not desirable anyway, since that implies the emulator is built with a fixed choice of fiducial cosmology). Instead, we could include $q_{\parallel}$ and $q_{\perp}$ as additional parameters in the emulator, but this would increase the emulation parameter space with potentially negative impacts on the overall accuracy of the emulator. For that reason we opt for emulating $P_{{\cal B}}(\boldsymbol{k}|z)$ without AP distortions and fold in their effect analytically in a separate step.

\begin{table*}
  \centering
  \caption{Bias contributions to the power spectrum at linear and one-loop order, which scale as $P_L$ and $P_L^2$, respectively.}
  \begin{tabular}{c|ccccccccccccccccccc}
    \hline
    ${\cal B}$ & $b_1^2$ & $b_1$ & $1$ & $b_1^2$ & $b_1\,b_2$ & $b_1\,\gamma_2$ & $b_1\,\gamma_{21}$ & $b_2^2$ & $b_2\,\gamma_2$ & $\gamma_2^2$ & $b_2$ & $\gamma_2$ & $\gamma_{21}$ & $c_0$ & $c_2$ & $c_4$ & $b_1^2\,c_{\rm nlo}$ & $b_1\,c_{\rm nlo}$ & $c_{\rm nlo}$ \\ \hline
    linear & \checkmark & \checkmark & \checkmark &  &  &  &  &  &  &  &  &  &  & \checkmark & \checkmark & \checkmark & \checkmark & \checkmark & \checkmark \\
    one-loop &  & \checkmark & \checkmark & \checkmark & \checkmark & \checkmark & \checkmark & \checkmark & \checkmark & \checkmark & \checkmark & \checkmark & \checkmark &  &  &  & & &   \\ \hline
  \end{tabular}
  \label{tab:emu.contributions}
\end{table*}

Another reason for not including the AP factors in the emulator itself is that we can also separately apply the effective damping function $W_{\infty}$ in the VDG model (see Sec.~\ref{sec:RSD_power}). As $W_{\infty}$ may depend on model parameters $\boldsymbol{\Theta}_{W}$ (in addition to $\boldsymbol{\Theta}_{\rm s}$ and $\boldsymbol{\Theta}_{\rm e}$), such as $a_{\rm vir}$, that do not factorise, we can thus further limit the total number of parameters whose dependence needs to be emulated. Therefore, given the emulated $P_{{\cal B}}(k',\mu'|z)$, we compute the final multipoles by performing the integration
\begin{equation}
  \label{eq:emu.Pell_wAP}
  \begin{split}
    P_{{\cal B},\ell}(k|z,\boldsymbol{\Theta}_{\rm s},\boldsymbol{\Theta}_{\rm e}) = &\ \frac{2\ell + 1}{2 q_{\perp}^2 q_{\parallel}} \int_{-1}^1 \mathrm{d}\mu\, {\cal L}_{\ell}(\mu)\, P'_{{\cal B}}(k',\mu'|z,\boldsymbol{\Theta}_{\rm s},\boldsymbol{\Theta}_{\rm e})\, \\ &\ \times\,W_{\infty}(k',\mu'|z,\boldsymbol{\Theta}_{\rm s},\boldsymbol{\Theta}_{\rm e},\boldsymbol{\Theta}_{W})\,,
  \end{split}
\end{equation}
where $W_{\infty} = 1$ in case of the EFT model and primes denote quantities in the fiducial cosmology. They are related to those in the true cosmology as follows \citep{Bal9605}:
\begin{align}
  k' &= \frac{k}{q_{\perp}} \left[1 + \mu^2 \left(F^{-2}-1\right)\right]^{1/2} \\
  \mu' &= \frac{\mu}{F} \left[1 + \mu^2 \left(F^{-2}-1\right)\right]^{-1/2}\,,
\end{align}
where $F \equiv q_{\parallel}/q_{\perp}$. We note that the dependence of $W_{\infty}$ on the shape and evolution parameters enters through the growth rate and the velocity dispersion, which is a functional of the linear power spectrum (see Eq.~\ref{eq:sigmav}). That means that as for $P_{{\cal B}}$ the cosmology and redshift dependence of the damping term can be fully captured by $\boldsymbol{\Theta}_{\rm s}$, the growth rate and $\sigma_{12}$.

The expression in Eq.~(\ref{eq:emu.Pell_wAP}) assumes that the power spectrum multipoles are estimated in infinitesimally thin shells $k$ and for a continuous range of $\mu$ values. Both of these assumptions are not satisfied for actual measurements, where the multipoles are computed from discrete grids in Fourier space and by averaging over all wave modes that fall into shells of given widths \citep[see e.g.][]{Sco1510}. As shown in \citet{TarNisBer1304} this leads to sizeable deviations from what Eq.~(\ref{eq:emu.Pell_wAP}) would predict, in particular for the higher-order multipoles and for $k$ values close to the fundamental frequency, $k_f$, of the grid used for the measurements. As described in more detail in Appendix~\ref{sec:discreteness}, one can account for these discreteness effects in the theoretical predictions, which we have implemented as an option in \emu. Up to scales $k \sim 60\,k_f$ this comes at no additional computational cost compared to the computation via Eq.~(\ref{eq:emu.Pell_wAP}).

\subsection{Emulated quantities}
\label{sec:emulated_quantities}

As is clear from the previous two sections, the task of our emulator is to provide the contributions $P_{{\cal B}}(\boldsymbol{k}|z)$. In total there are 17 different contributions to take into account, summarised in Table~\ref{tab:emu.contributions}. For ${\cal B} = b_1$ and ${\cal B}=1$ we combine linear and one-loop terms, while for ${\cal B} = b_1^2$ we emulate them separately, allowing our emulator to also provide the IR resummed linear power spectrum, which can be useful for other computations. The stochastic contributions (involving the noise parameters $N^P_0$, $N^P_{2,0}$ and $N^P_{2,2}$) do not appear in this list because they do not depend on any cosmological quantities themselves and since we apply AP distortions and the finger-of-god damping separately (see Eq.~\ref{eq:emu.Pell_wAP}), we can include them exactly.

\subsubsection{Projection into multipoles and reconstruction of anisotropic power spectrum}
\label{sec:emu_reconstruction}

In order to avoid having to emulate the full two-dimensional dependence of $P_{{\cal B}}$ on $k$ and $\mu$, we project the angular dependence into multipoles,
\begin{equation}
  \label{eq:emu.Pell}
  P_{{\cal B},\ell}(k) = \frac{2\ell+1}{2} \int_{-1}^1 \mathrm{d}\mu\, {\cal L}_{\ell}(\mu)\, P_{{\cal B}}(k,\mu)\,,
\end{equation}
and emulate only the monopole, quadrupole and hexadecapole, from which we reconstruct the 2D power spectra using the Legendre expansion. This procedure would be exact if all of the contributions contain powers of, at most, $\mu^4$. However, the redshift-space galaxy power spectrum contains terms up to $\mu^8$ and, moreover, the dependence of the IR damping factor on $\mu^2$ leads to non-zero multipoles for all even $\ell$. Our reconstruction using only information up to the hexadecapole therefore introduces an error that becomes more relevant for the higher-order multipoles computed through Eq.~(\ref{eq:emu.Pell_wAP}). In order to approximately correct for that we include the $\ell=6$ multipole $P_{{\cal B},6}$, evaluated at fixed redshift $z=1$ and for a fixed set of $\Lambda$CDM cosmological parameters taken from the \emph{Planck} TT,TE,EE+lowE+lensing constraints \citep{Planck2018} in the Legendre expansion:
\begin{equation}
  \label{eq:emu.reconstruction}
  \begin{split}
    P_{{\cal B}}(\boldsymbol{k}|z,\boldsymbol{\Theta}_{\rm s},\boldsymbol{\Theta}_{\rm e}) \approx &\ \sum_{\ell = 0}^2 P_{{\cal B},2\ell}(k|z,\boldsymbol{\Theta}_{\rm s},\boldsymbol{\Theta}_{\rm e})\, {\cal L}_{2\ell}(\mu) \\ &\ \hspace{-7.1em} + P_{{\cal B},6}\left\{k|f(z,\boldsymbol{\Theta}_{\rm s},\boldsymbol{\Theta}_{\rm e}), P_L\big(k|\boldsymbol{\Theta}^{\rm Planck}_{\rm s},\sigma_{12}(z,\boldsymbol{\Theta}_{\rm s},\boldsymbol{\Theta}_{\rm e})\big)\right\}\, {\cal L}_6(\mu)\,.
  \end{split}
\end{equation}
Our notation highlights that the fixed Planck values only enter through the shape parameters affecting the linear power spectrum, while the dependence of the growth rate and $\sigma_{12}$ on the shape and evolution parameters is correctly accounted for. This is achieved by splitting up each $P_{{\cal B},6}$ into contributions with different powers of $f$ and scaling their amplitude as follows:
\begin{equation}
  \label{eq:emu.P6_scaling}
  \left.P_{{\cal B},6}\right|_{\rm Planck} \;\rightarrow\; \left(\frac{\sigma_{12}(z,\boldsymbol{\Theta}_{\rm s},\boldsymbol{\Theta}_{\rm e})}{\sigma_{12}\left(z=1,\boldsymbol{\Theta}^{\rm Planck}_{\rm s},\boldsymbol{\Theta}^{\rm Planck}_{\rm e}\right)}\right)^{2L}\,\left.P_{{\cal B},6}\right|_{\rm Planck}\,,
\end{equation}
where $L=1$ for all terms indicated as linear in Table~\ref{tab:emu.contributions} and $L=2$ for the one-loop terms.\footnote{In this way only the dependence of the IR damping term on $f$ and $\sigma_{12}$ is not correctly taken into account, but instead computed for $z=1$ and at the fixed Planck cosmological parameters.} In Sec.~\ref{sec:reconstruction} we quantify the inaccuracies introduced by Eq.~(\ref{eq:emu.reconstruction}) and demonstrate that they are negligible.

\subsubsection{Constructing ratios with the linear power spectrum}

The amplitude of $P_{{\cal B},\ell}$ can be subject to large variations over the full range of values that the emulated parameters can assume. As that makes the emulation more difficult, we instead emulate the ratios, $\beta_{{\cal B},\ell}$, of $P_{{\cal B},\ell}$ and the linear power spectrum (excluding IR resummation), which significantly reduces the dynamical range of the relevant quantities. After emulating the ratios we then need to multiply again by the linear power spectrum, for which we build a separate emulator. However, this emulator can be constructed over the shape parameters alone, as shown by the following (exact) computation:
\begin{align}
  \label{eq:emu.PBell_from_ratio}
  P_{{\cal B},\ell}(k|z,\boldsymbol{\Theta}_{\rm s},\boldsymbol{\Theta}_{\rm e}) &= \beta_{{\cal B},\ell}(k|\boldsymbol{\Theta}_{\rm s},\sigma_{12},f)\, P_L(k|\boldsymbol{\Theta}_{\rm s},\sigma_{12}) \nonumber \\
                                                                                 &= \beta_{{\cal B},\ell}(k|\boldsymbol{\Theta}_{\rm s},\sigma_{12},f)\, P_L(k|z=1,\boldsymbol{\Theta}_{\rm s},\boldsymbol{\Theta}_{\rm e}^{\rm fixed}) \nonumber \\
  &\times\, \left(\frac{\sigma_{12}}{\sigma_{12}\left(z=1,\boldsymbol{\Theta}_{\rm s},\boldsymbol{\Theta}_{\rm e}^{\rm fixed}\right)}\right)^2\,,
\end{align}
where we set $\sigma_{12} = \sigma_{12}(z,\boldsymbol{\Theta}_{\rm s},\boldsymbol{\Theta}_{\rm e})$ and used that the dependence on $\sigma_{12}$ can be factored out of the linear power spectrum by evaluating $P_L$ at fixed redshift and evolution parameters and rescale the amplitude accordingly.\footnote{The fixed values for $z$ and the evolution parameters are not essential, but we used $z=1$, $h = 0.695$, $A_s = 2.2078559$ and all other potential evolution parameters set to zero.} The value of $\sigma_{12}\left(z=1,\boldsymbol{\Theta}_{\rm s},\boldsymbol{\Theta}_{\rm e}^{\rm fixed}\right)$ can be obtained as an integral over $P_L(k|z=1,\boldsymbol{\Theta}_{\rm s},\boldsymbol{\Theta}_{\rm e}^{\rm fixed})$, but instead we find it is more efficient to include an emulator for the former as a function of only $\boldsymbol{\Theta}_{\rm s}$. As noted in Sec.~\ref{sec:emu_AP+vir} the finger-of-god damping factor in the VDG model depends on the velocity dispersion and a computation similar to that in Eq.~(\ref{eq:emu.PBell_from_ratio}) shows that it is sufficient to predict the value of $\sigma_v(z=1,\boldsymbol{\Theta}_{\rm s},\boldsymbol{\Theta}_{\rm e}^{\rm fixed})$, which we also emulate as a function of the shape parameters.

\subsubsection{Summary}

In summary, we thus require an emulation of the following quantities:
\begin{enumerate}
  \setlength\itemsep{0.8em}
\item $\beta_{{\cal B},\ell}(k|\boldsymbol{\Theta}_{\rm s},\sigma_{12},f)$ for $\ell = 0,2,4$ and each ${\cal B}$ from Table~\ref{tab:emu.contributions}\,,
\item $P_L\left(k|z=1,\boldsymbol{\Theta}_{\rm s},\boldsymbol{\Theta}_{\rm e}^{\rm fixed}\right)$\,,
\item $\sigma_{12}\left(z=1,\boldsymbol{\Theta}_{\rm s},\boldsymbol{\Theta}_{\rm e}^{\rm fixed}\right)$\,,
\item $\sigma_{v}\left(z=1,\boldsymbol{\Theta}_{\rm s},\boldsymbol{\Theta}_{\rm e}^{\rm fixed}\right)$\,.
\end{enumerate}
We evaluate the ratios and $P_L$ on a range of scales extending from $7 \times 10^{-4}\,\invMpc$ to $0.35\,\invMpc$, using a total number of 106 points, chosen such that they provide a dense sampling on scales relevant for the BAO wiggles.

\vspace*{-0.5em}
\subsection{Parameter space and training process}
\label{sec:training}

\begin{table}
  \centering
  \caption{Emulator parameter space and the supported range of values.}
  \begin{tabular}{ccc}
    \hline
    Parameter & Min. emulator range & Max. emulator range \\ \hline
    $\omega_b$ & $0.0205$ & $0.02415$ \\
    $\omega_c$ & $0.085$ & $0.155$ \\
    $n_s$ & $0.92$ & $1.01$ \\ \hline
    $\sigma_{12}$ & $0.2$ & $1.0$ \\
    $f$ & $0.5$ & $1.05$ \\ \hline
  \end{tabular}
  \label{tab:emu_ranges}
\end{table}

\begin{figure}
  \centering
  \includegraphics{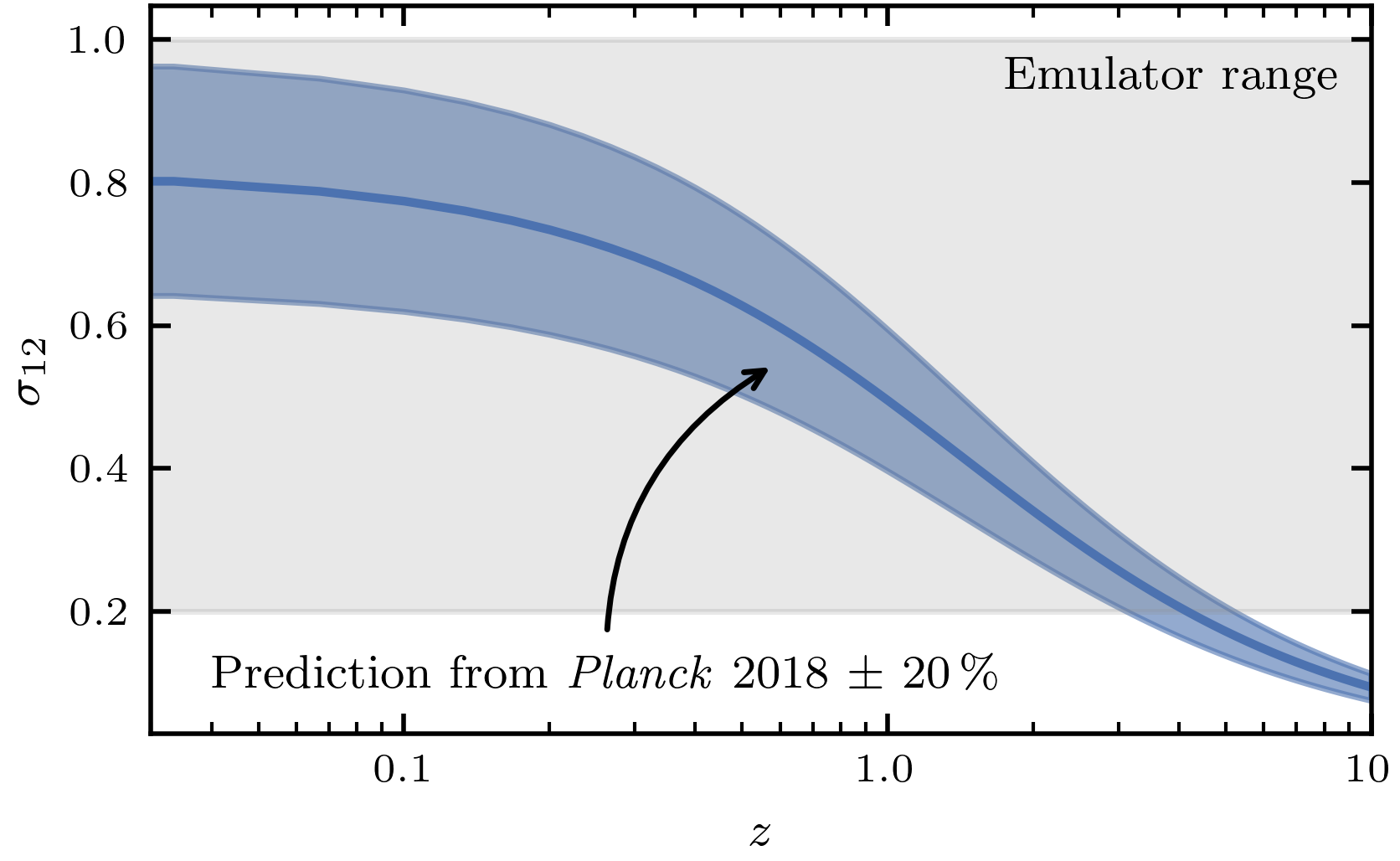}
  \vspace*{-1em}
  \caption{Prediction of $\sigma_{12}$ as a function of redshift using the \emph{Planck} 2018 TT,TE,EE+lowE+lensing best-fit cosmological parameters. The blue shaded band represents a 20\,\% variation around that prediction, while for comparison the supported range of values for $\sigma_{12}$ in our emulator \emu\ is shown as the grey shaded area.}
  \label{fig:emu_s12_of_z}
\end{figure}

\subsubsection{Parameter ranges}

Our emulator is constructed for a total of five parameters: in addition to $f$ and $\sigma_{12}$, we consider the three shape parameters $\omega_b$, $\omega_c$, and $n_s$. Each of them is allowed to vary within the ranges given in Table~\ref{tab:emu_ranges}, which for the latter three were chosen to span roughly a 12, 30 and 11$\sigma$ interval around the \emph{Planck} 2018 best-fit values, respectively. The growth rate and $\sigma_{12}$ capture the dependence on redshift and an arbitrary set of evolution parameters\footnote{In our current release of \emu\ we allow to specify values for $h$, $A_s$, $\omega_K$, $w_0$, and $w_a$.}, and therefore require a more generous support. Nonetheless, the limits on $f$ and $\sigma_{12}$ impose restrictions on the range of redshifts for which our emulator can be used. In case of the growth rate the ranges can accommodate any redshift for $\omega_c \gtrsim 0.107$, while for the most extreme values of the allowed shape parameters the lower boundary imposes the limitation $z \gtrsim 0.1$. Tighter restrictions on the supported redshifts come from the range of $\sigma_{12}$ and to demonstrate that we compare them in Fig.~\ref{fig:emu_s12_of_z} against the \emph{Planck} prediction of $\sigma_{12}$ as a function of redshift using the best-fit cosmological parameter values from \cite{Planck2018}. When exploring cosmological parameters using a large-scale structure likelihood function, they will give rise to values of $\sigma_{12}$ that are typically close (within $\sim 10\,\%$) to the \emph{Planck} prediction, while the 1-$\sigma$ uncertainty on $\sigma_{12}$ is of the order $5\,\%$ for constraints from the BOSS galaxy survey \citep{SemSanPez2206}. To account for these uncertainties we plot a 20\,\% error band around the \emph{Planck} prediction in Fig.~\ref{fig:emu_s12_of_z} and expect that any sampled cosmological parameters will correspond to $\sigma_{12}$ values falling roughly into that range. We see that this band leaves the $\sigma_{12}$ range of our emulator at a redshift $z \sim 3$, which means that \emu\ is no longer guaranteed to provide accurate predictions beyond that redshift.

\vspace*{-0.5em}
\subsubsection{Generation of training data sets}

We generate two separate training sets, one spanning the full parameter space intended for the ratios $\beta_{{\cal B},\ell}$, and another covering only the shape parameters for the remaining quantities. Both training sets are built by drawing a number of samples from a Latin Hypercube, using 1500 and 750 samples, respectively. In order to obtain an optimal coverage of the parameter spaces we repeat the sampling step 10,000 times and pick the set which maximises the minimum (Euclidean) distance between any two of its points. We then evaluate all of the model ingredients using a numerical integrator and starting from \texttt{CAMB}-generated linear input power spectra. Before training the emulator we perform one additional pre-processing step, in which we further reduce the dynamical range of the training data by taking the logarithm and in which we normalise each component, such that it has zero mean and unit variance over the full set of samples.

\subsubsection{Gaussian process emulation}

For the actual emulation step we employ Gaussian Processes (GP), whose properties are discussed in detail in \citet{RasWil06}. The crucial ingredient in a GP model is the kernel function $K(\boldsymbol{x},\boldsymbol{x}')$, which describes the covariance between two points $\boldsymbol{x}$ and $\boldsymbol{x}'$ of the training set. Due to a lack of knowledge about the precise functional form of this covariance we generated emulators with different combinations of commonly used kernel functions in the literature. After comparing their performance we settled on the following kernel function
\begin{equation}
  \label{eq:emu.kernel}
  K(\boldsymbol{x},\boldsymbol{x}') = K_{\rm exp}(\boldsymbol{x},\boldsymbol{x}') + K_{3/2}(\boldsymbol{x},\boldsymbol{x}')\,,
\end{equation}
which is a combination of a squared exponential kernel,
\begin{equation}
  \label{eq:emu.kernel_exp}
  K_{\rm exp}(\boldsymbol{x},\boldsymbol{x}') = \exp{\left(-\frac{r^2}{2}\right)}\,,
\end{equation}
and a Mat\'ern kernel of degree $\nu = 3/2$,
\begin{equation}
  \label{eq:eme.kernel_matern}
  K_{3/2}(\boldsymbol{x},\boldsymbol{x}') = \left(1 + \sqrt{3} r\right)\,\exp{\left(-\sqrt{3} r\right)}\,,
\end{equation}
where $r^2 = \sum_{i=1}^d \left(x_i - x'_i\right)^2/l_i^2$ and $d$ is the dimension of the parameter space. The quantities $\boldsymbol{l}$ represent so-called hyperparameters, which characterise the length scales of typical features in the training data and they can differ between the two kernel functions. The values of these hyperparameters are optimised by maximising the log-likelihood of our GP models with respect to the training data. We implement this procedure using the publically available package \texttt{GPy} and repeat the optimisation step five times with different random initialisations, selecting for each emulator the parameter set that provides the largest log-likelihood.

\subsection{Computational performance}
\label{sec:time}

In this section we measure the execution times of \emu\ for the two different RSD models, each for a different number of multipoles and number of scales. The computational performance will of course depend on the given platform, so the reader should keep in mind that all measurements reported here are based on a laptop equipped with an Apple M1 Pro processor, using one CPU (up to 3.22 Ghz) and a single thread.

\begin{figure}
  \centering
  \includegraphics{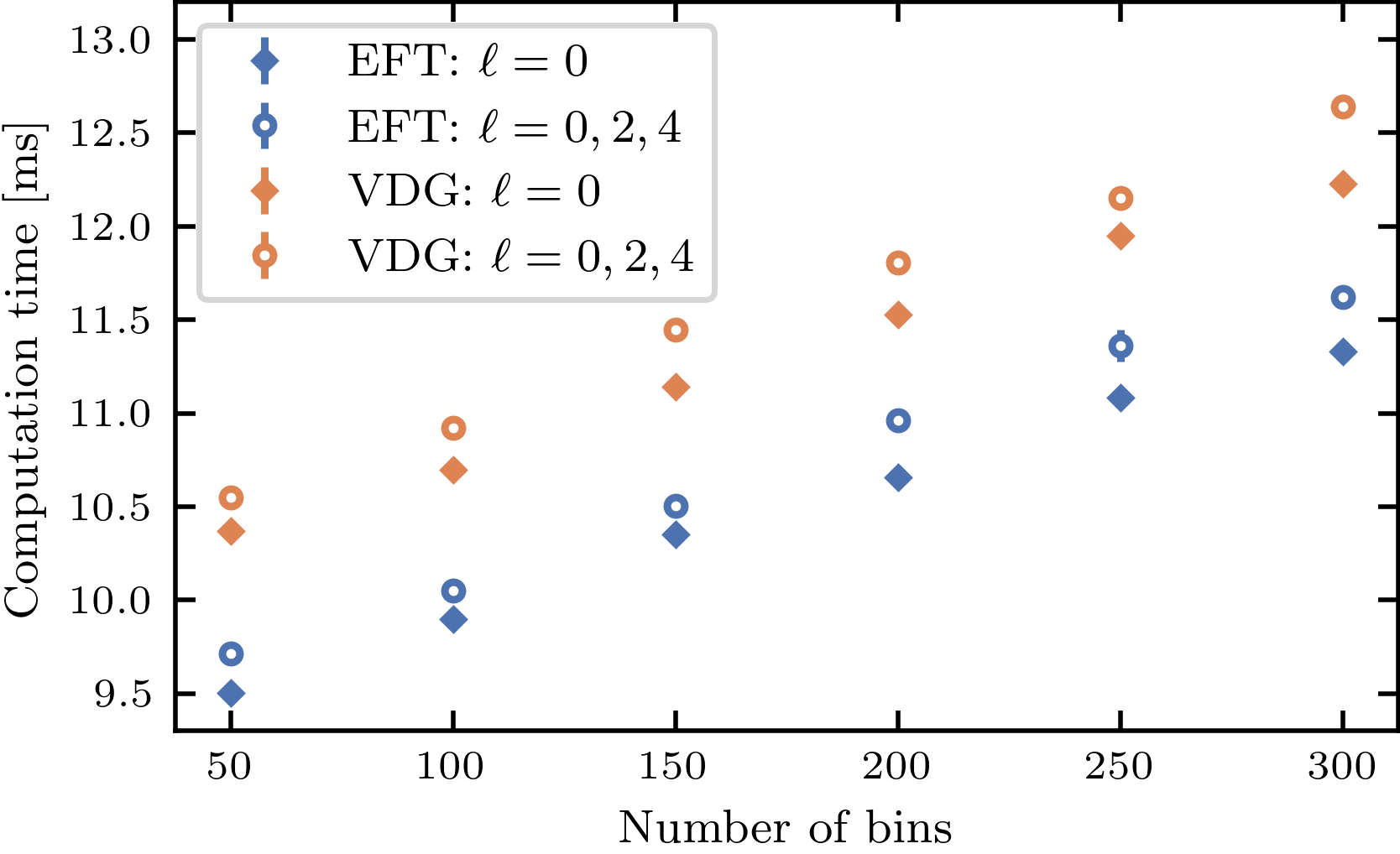}
  \caption{Computation time as a function of the provided number of bins for the EFT and VDG models, either for the monopole alone, or all three multipoles.}
  \label{fig:computation_time}
\end{figure}

We measure the execution times using the function \texttt{perf\_counter} from \texttt{Python}'s \texttt{time} module and in order to reduce uncertainties, we repeat the prediction of the power spectrum multipoles $N_{\rm exec}$ times for values of $N_{\rm exec}$ ranging from 1 to 35. We then fit a straight line to the resulting measurements as a function of $N_{\rm exec}$, such that the slope of that line provides a robust estimate (and uncertainty) of the execution time per call. In Fig.~\ref{fig:computation_time} we plot these estimates for a varying number of scales spaced logarithmically between $k_{\rm min} = 0.001\,\invMpc$ and $k_{\rm max} = 0.35\,\invMpc$ for the EFT model in blue and the VDG model in orange. Filled symbols in each case correspond to the prediction of only the monopole, while open symbols also include the computation of the quadrupole and hexadecapole. We see that the execution times range from $\sim 9.5\,\mathrm{ms}$ to $\sim 11.5\,\mathrm{ms}$ for 50 to 300 bins in case of the EFT model. The VDG model takes on average about $1\,\mathrm{ms}$ longer, due to the integration over $\mu$ also involving the evaluation of the effective damping function $W_{\infty}$. On the other hand, we see little difference between a prediction of just the monopole, or all three multipoles, since for the reconstruction of the anisotropic power spectrum the emulator of all three multipoles need to be called in any case.

Considering that the execution time for other public codes, such as \texttt{CLASS-PT} \citep{ChuIvaPhi2009} or \texttt{PyBird} \citep{dAmSenZha2101} is of the order $\sim 1\,\mathrm{s}$ (based on a timing estimate given by the authors of the former code, but using more than one CPU), \emu\ achieves a speed-up of at least two orders of magnitude.

%%%%%%%%%%%%%%%%%%%%%%%%%%%%%%%%%%%%%%%%%%%%%%%%%%%%%%%%%%%%%%%%%%%%%%%%%%%%%%%%%%%%%%%%%%%%%%%%%%%%%%%%%%%%%%%%%%%%%%%%%%%%%%%%%%%%%%%%%%%%%%%%%%%%%%
\section{Validation of the emulator}
\label{sec:validation}

In this section we perform a number of mock analyses based on synthetic data sets at multiple redshifts using statistical uncertainties corresponding to a volume ten times larger than that expected for the \Euclid\ galaxy survey. These analyses not only allow us to determine the relative uncertainties introduced by the emulator, they also --- and this is of much greater relevance --- let us test how these uncertainties propagate to the posteriors of cosmological and galaxy bias parameters. 

\subsection{Generation of synthetic validation data}
\label{sec:synthetic_data}

\subsubsection{$\Lambda$CDM validation set}
\label{sec:LCDM_validation_set}

\begin{table}
  \centering
  \caption{Parameters included in the $\Lambda$CDM validation set and their minimum and maximum values. The parameter $a_{\rm vir}$ is only included in the validation set for the VDG model and its minimum and maximum values are given in units of $\hMpc$.}
  \begin{tabular}{ccc}
    \hline
    Parameter & Minimum & Maximum \\
    \hline
    $\omega_b$  & 0.02100 & 0.02365 \\
    $\omega_c$  & 0.095 & 0.145 \\
    $n_s$       & 0.93  & 1.00 \\
    $h$        & 0.55 & 0.85 \\
    $A_s$      & 0.8  & 3.0 \\
    $a_{\rm vir}$ & 0.0 & 8.0 \\
    \hline
  \end{tabular}
  \label{tab:LCDM_validation_ranges}
\end{table}
\begin{figure*}
  \centering
  \includegraphics{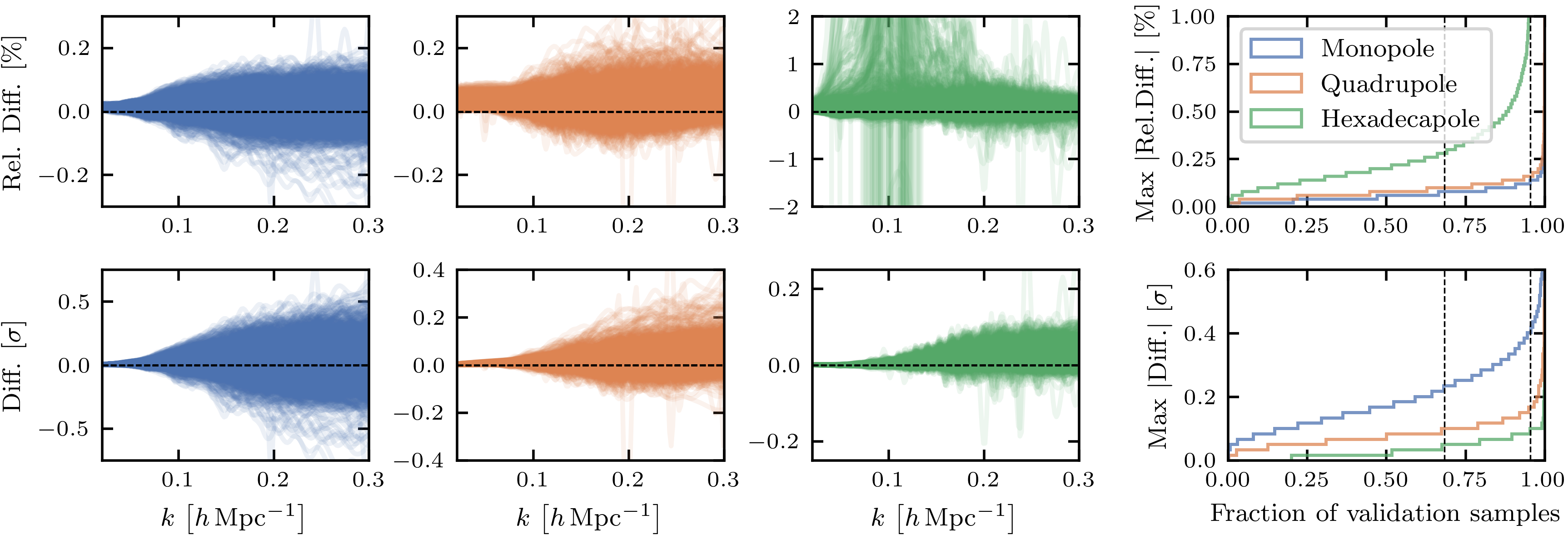}
  \caption{Inaccuracies of the emulated multipoles as a function of scale for the EFT model at $z=0.9$. Differences are shown in percent (upper panels) and in units of the standard deviation of our synthetic data set (covering 10 times the volume of a \Euclid\ redshift shell, see Sec.~\ref{sec:LCDM_validation_set}; lower panels) for all cosmologies of the validation set and using a combination of bias parameters from a random point in each chain. The fourth panel shows the cumulative histogram of the maximum absolute differences over the full range of scales with the two vertical dashed lines indicating $68\,\%$ and $95\,\%$ of the validation samples.}
  \label{fig:emu_comparison_rel_abs_diff}
\end{figure*}

Our main validation sample consists of a set of flat $\Lambda$CDM cosmologies covering the parameters $\omega_b$, $\omega_c$, $n_s$, $h$, and $A_s$ for the EFT model, while for the VDG model we also include $a_{\rm vir}$. Each validation set is generated by drawing 1500 random points within these five-, or six-dimensional parameter spaces, satisfying only the minimum and maximum values given in Table~\ref{tab:LCDM_validation_ranges}. For each point in the validation set we compute the power spectrum multipoles using the exact model at the four redshifts $z = 0.9$, $1.2$, $1.5$, and $1.8$, and making the following assumptions for the galaxy bias parameters: we fix the value of the linear bias using $b_1(z) = \sqrt{1+z}$, which provides a reasonable estimate of the bias of $\mathrm{H}{\alpha}$ galaxies to be selected by \Euclid\ \citep{RasAmaAme0810,diPAmeBra1201}, whereas for $\gamma_2$ we impose the excursion-set relation of \citet{SheChaSco1304, EggScoCro2011}. We then determine the values of $b_2$ and $\gamma_{21}$ according to the peak-background-split and coevolution relations from \citet{LazWagBal1602} and \citet{EggScoSmi1906}, respectively, as functions of $b_1$ and $\gamma_2$, and set all other counterterm and stochastic model parameters to zero. Each of the synthetic multipoles accounts for Alcock-Paczynski distortions in an exact manner (i.e., not via the approximation discussed in Sec.~\ref{sec:emu_reconstruction}), based on a fiducial flat $\Lambda$CDM cosmology at the same redshifts as the predictions and with fixed parameters $h = 0.67$ and $\omega_m = \omega_c + \omega_b = 0.1432$.

In a second step we construct statistical uncertainties for these synthetic measurements by adopting Gaussian covariances matrices, which are computed from the synthetic multipoles at each point in the validation set using the expressions given, for instance, in \citet{GriSanSal1604}. In order to relate these uncertainties to the characteristics of the \Euclid\ survey, we further assume tracer densities that match the number densities of $\mathrm{H}\alpha$ galaxies in the \Euclid\ Flagship I
mock catalogues (see Table~\ref{tab:validation_params}), as well as \Euclid-like volumes. The latter are derived from redshift shells of width $\Delta z = 0.2$, centred on the respective redshifts for which the power spectrum multipoles have been evaluated, and covering a sky area of $15,000\,\mathrm{deg}^2$. To increase the stringency of our validation tests, we then multiply these volumes by a factor of 10, which is equivalent with a reduction of the statistical uncertainties by a factor of
$\sim 3$. 

We stress that the various choices in generating this synthetic data set are not meant to provide power spectrum measurements that closely mimic those of any real galaxy samples. Nonetheless, we expect the relative uncertainties, $\sigma_{\ell}/P_{\ell}$, to be well representative (apart from the stringency volume factor) for \Euclid, such that we can make a meaningful assessment of the performance of our emulator.

\subsubsection{Synthetic measurements for fixed set of parameters}
\label{sec:fixed_cosmology}

For a further validation test we generate one more set of synthetic power spectrum multipoles at fixed cosmological and bias parameters. We compute the power spectrum multipoles from the exact model (in this case only for the EFT) at the same four redshifts as before, but using the bias parameter values for $b_1$, $b_2$, $\gamma_{21}$, $c_0$, $c_2$, $c_4$, $c_{\rm nlo}$ and $N^P_0$ given in Table~\ref{tab:validation_params}. The value for $\gamma_2$ is again fixed in terms of the excursion set relation, while the cosmological parameters at all four redshifts are set to $h = 0.67$, $\omega_c = 0.1212$, $\omega_b = 0.021996$, $n_s = 0.96$ and $A_s = 2.11065$. As described in Sec.~\ref{sec:LCDM_validation_set} we then derive statistical uncertainties for these synthetic measurements in the Gaussian approximation, using the number densities specified in Table~\ref{tab:validation_params} and volumes that correspond to the same redshift shells as above, including the stringency factor of 10.

\subsection{Results across $\Lambda$CDM validation set}
\label{sec:LCDM_validation}

Our goal is to quantify the impact of the emulation inaccuracies on mean parameter values and their uncertainties, when performing a full likelihood exploration using all three galaxy power spectrum multipoles. To that end we use the synthetic data vectors and covariance matrices described in Sec.~\ref{sec:LCDM_validation_set} and for each combination of cosmological parameters in the validation set we run two Monte Carlo Markov chains\footnote{These chains are run with \texttt{emcee} \citep{ForHogLan1303}, using 32 walkers and for a number of steps at least ten times the maximum auto-correlation length for all varied parameters. The chains are then post-processed with \texttt{getdist} \citep{Lew1910} in order to extract posteriors and related statistics.} (MCMC), one with the exact theory model, the other using the emulator predictions. In those chains we keep the cosmological parameters fixed, while varying a set of seven bias parameters: $b_1$, $b_2$, $\gamma_{21}$, the constant shot noise $N_0$, as well as the three counterterm parameters $c_0$, $c_2$, and $c_4$; $\gamma_2$ is fixed in terms of the aforementioned excursion set relation as a function of $b_1$. We pick a maximum wavemode of $0.3\,\hinvMpc$ for all three multipoles in this analysis, which means that any significant deviations between the true and emulated models up to that scale can lead to shifts between the posterior mean values recovered from the two chains, as well as to  differences in the credible regions.

\begin{figure*}
  \centering
  \includegraphics{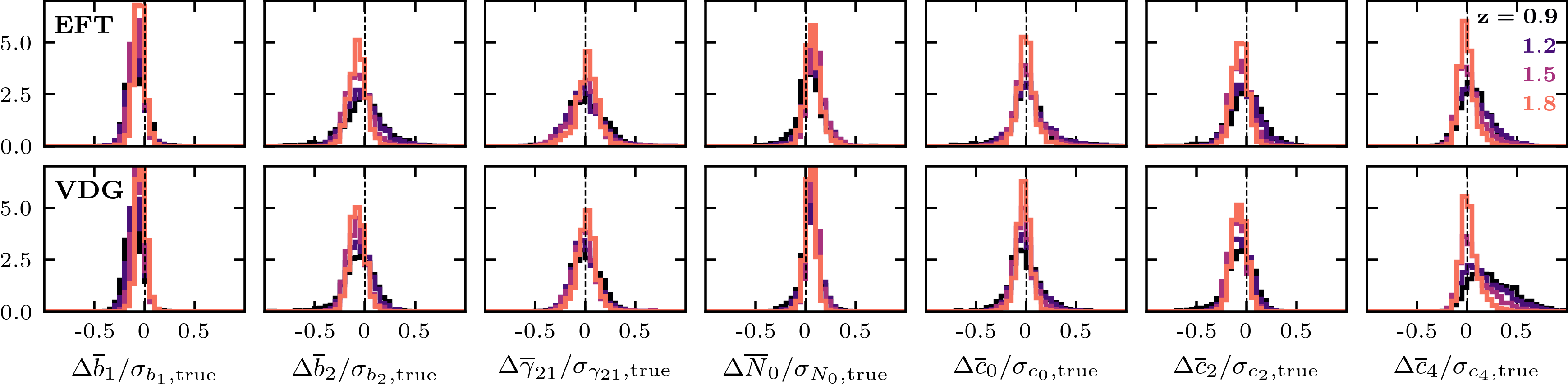}
  %\vspace*{-10mm}
  \caption{Distributions of the differences between the posterior mean values obtained from running MCMC with the exact and emulated model predictions. Chains were run for synthetic data including the monopole, quadrupole and hexadecapole up to $k_{\rm max} = 0.3\,\hinvMpc$ with uncertainties corresponding to expected errors for the \Euclid\ survey (see Sec.~\ref{sec:LCDM_validation_set}), but with ten times larger volumes. Each column shows the distribution for a different parameter varied in the chain in units of the respective standard deviation extracted from the chain using the true model, while the two different rows correspond to the different RSD models. The transition from dark to bright colours indicates increasing redshifts.}
  \label{fig:emu_comparison_means}
\end{figure*}

\subsubsection{Relative inaccuracies}
\label{sec:relative_inaccuracies}

\begin{figure}
  \centering
  \includegraphics{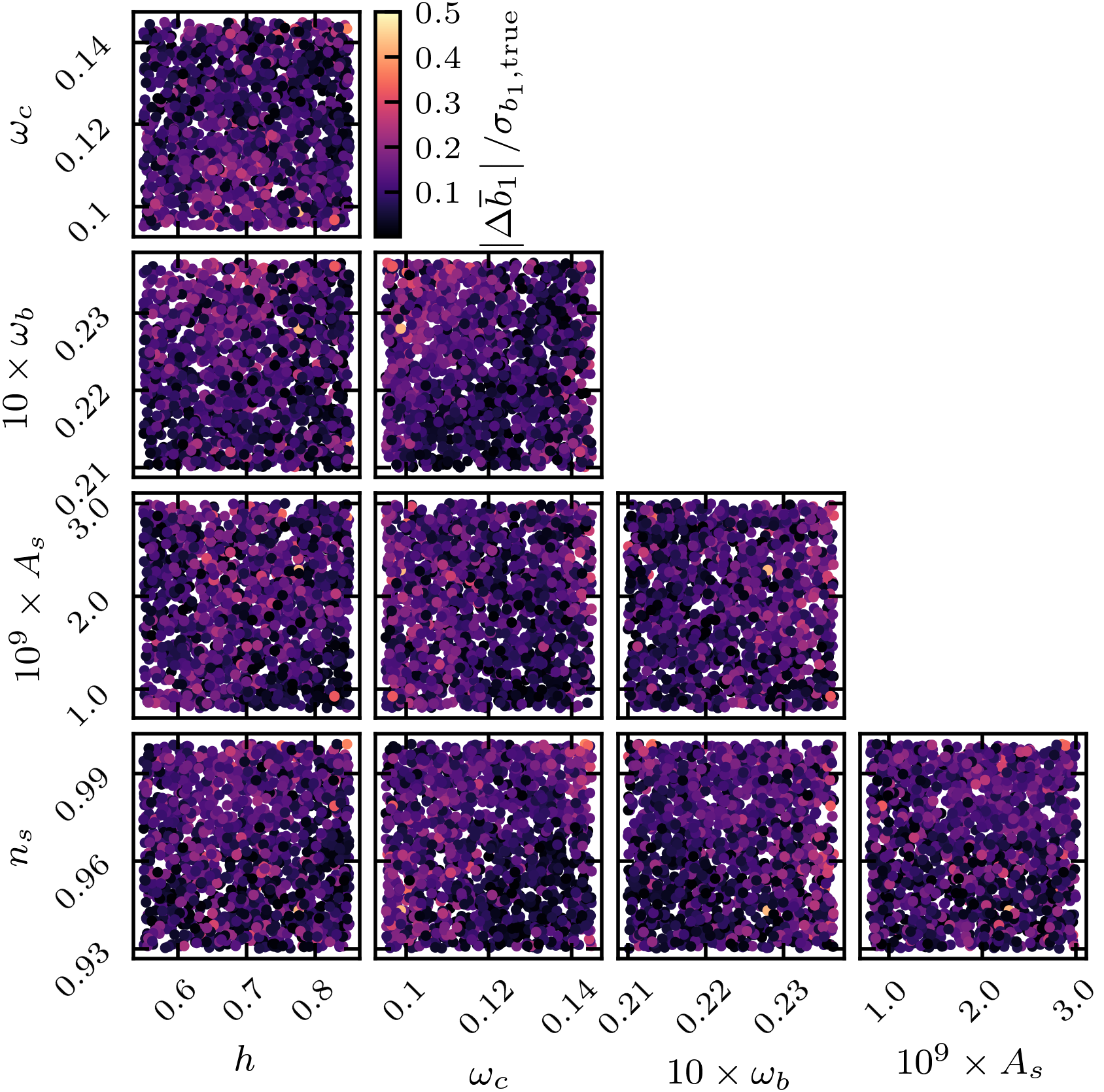}
  \vspace*{-1em}
  \caption{Shifts in the posterior means of $b_1$ between the true and emulated EFT model at $z=0.9$. Each panel shows a scatter plot of all validation samples, projected into different parameter planes.}
  \label{fig:EFT_diff_means_z0p9_b1}
\end{figure}
\begin{figure}
  \centering
  \includegraphics{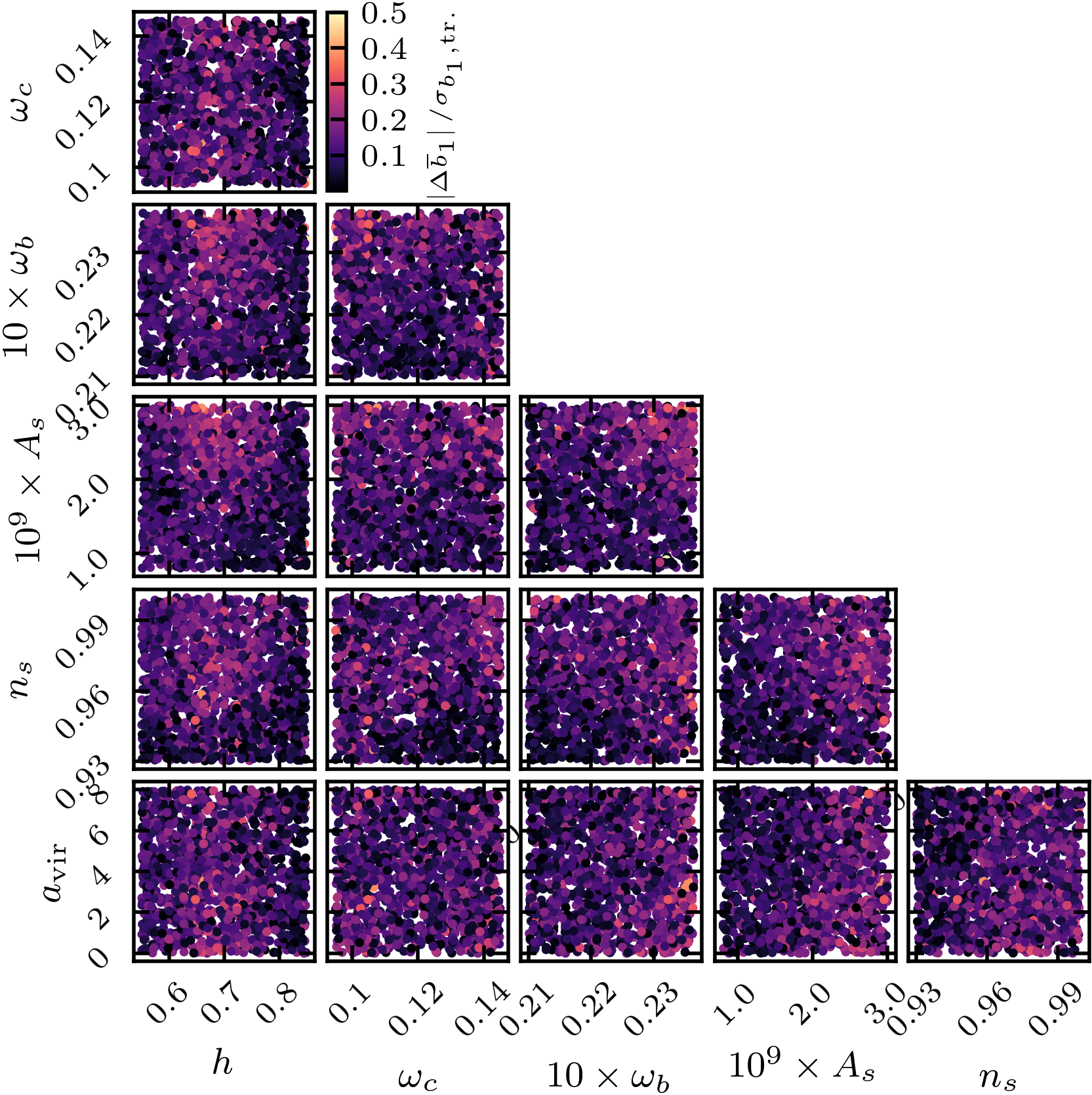}
  \caption{Same as Fig.~\ref{fig:EFT_diff_means_z0p9_b1}, but for the VDG model at $z=0.9$.}
  \label{fig:VIR_diff_means_z0p9_b1}
\end{figure}

Before identifying the effects on the parameter constraints, let us first consider the relative differences between the exact model and the emulator as a function of scale. These are shown exemplarily for the EFT model at $z=0.9$ in the upper panels of Fig.~\ref{fig:emu_comparison_rel_abs_diff}, where each line has been evaluated for a different point in the validation set and a randomly selected point from the chain over bias parameters that was run for the exact model. We see that the relative differences for the monopole (blue) and quadrupole (orange) grow between $k = 0.1\,\hinvMpc$ and $0.2\,\hinvMpc$, after which they saturate and generally stay below the $0.2\,\%$ threshold. This can be seen more clearly in the cumulative histogram in the fourth panel, which plots the maximum of the absolute relative difference over all scales and shows that there is only a vanishingly small fraction of cosmologies in the validation data sets that gives rise to discrepancies larger than $0.2\,\%$. In fact, we find that $68\,\%$ of all validation samples have maximum uncertainties smaller than $0.08\,\%$ and $0.1\,\%$ for the monopole and quadrupole, respectively. The situation is slightly worse for the hexadecapole, where for the same fraction of validation samples the maximum differences are only below $0.3\,\%$, but this is mostly due to the hexadecapole crossing zero for many of the tested cosmologies. It is more meaningful to plot these differences in units of some estimate of the measurement uncertainties, and for the results in the lower panels of Fig.~\ref{fig:emu_comparison_rel_abs_diff} we have picked the standard deviations taken from the covariance matrices discussed in Sec.~\ref{sec:LCDM_validation_set}. We now obtain the reversed picture: due to the monopole having the smallest uncertainties, its differences appear larger than for both, the quadrupole and hexadecapole, so it will dominate the impact on any parameter posteriors. However, $68\,\%$ of the validation samples still have a maximum difference smaller than $0.24\,\sigma$ at 10 times the \Euclid\ volume for this redshift shell. The analogous results for the VDG model and the other redshifts are qualitatively very similar --- we only note that the differences in units of the measurement uncertainties decrease with increasing redshift because of the decreasing tracer number densities and thus a larger contribution from shot noise, see Appendix~\ref{sec:app_cumulative_histograms}.

\subsubsection{Shifts in posterior means}
\label{sec:shifts_posterior_means}

\begin{figure*}
  \centering
  \includegraphics{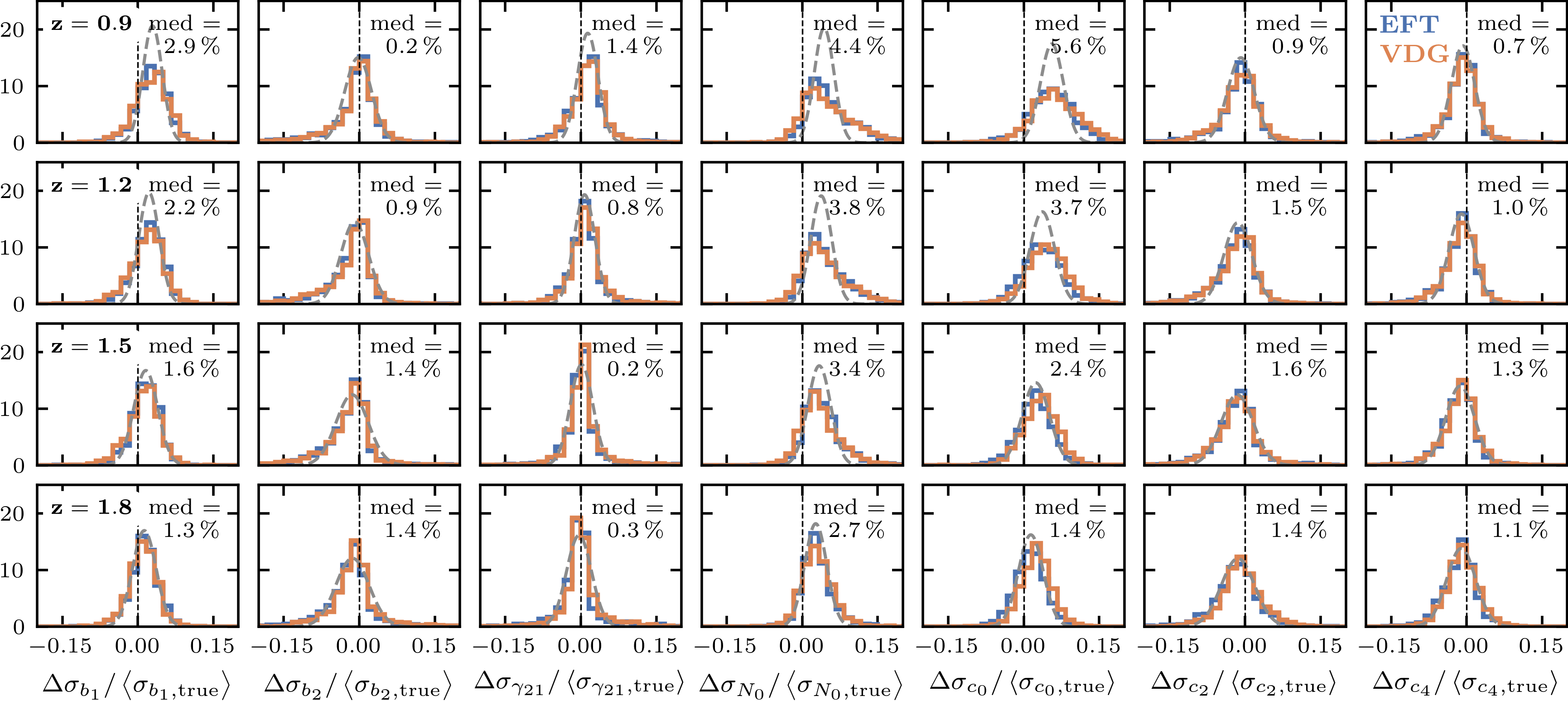}
  %\vspace*{-10mm}
  \caption{Histograms of the differences in the $68\,\%$ credible regions between the true and emulated models, normalised by the average 1-$\sigma$ constraints obtained from the true model of all validation samples, $\left<\sigma_{X,\mathrm{true}}\right>$. Each column depicts a different parameter varied in the chains, while each row shows a different redshift of the synthetic data set; the two different colours correspond to the two RSD models. The grey dashed lines are Gaussian distributions, centred on the median value of the EFT histograms, and represent an estimate of the spread due to sampling noise alone (see text for details).}
  \label{fig:emu_comparison_stds}
\end{figure*}

In order to study the impact of these inaccuracies on the parameter posteriors, we extract the 1d-marginalised posterior means for each of the seven bias parameters from the two chains and compute their difference in units of the 1-$\sigma$ parameter uncertainty obtained from the chain using the exact model, $\sigma_{X,\mathrm{true}}$ (with $X$ denoting any of the seven bias parameters). Fig.~\ref{fig:emu_comparison_means} shows the resulting histograms over the full validation set for all four redshifts and both of the RSD models. We clearly see that none of these cases produces any significant shifts in the varied parameters. More precisely, the means of the distributions stay consistently below $0.1\,\sigma_{X,\mathrm{true}}$, while the standard deviations reach a maximum of $0.2\,\sigma_{X,\mathrm{true}}$, showing that for the majority of the validation cosmologies we recover the true posterior means with high accuracy, while even the largest shifts remain negligible for the nominal \Euclid\ volume. The parameters $b_1$ and $N_0$ generally display the smallest shifts, since they already get well constrained by the large-scale power spectrum, where the emulation inaccuracies are smallest. The higher-order bias parameters $b_2$ and $\gamma_{21}$, as well as the counterterm parameters, on the other hand, are mostly constrained from the non-linear regime, such that they are more susceptible to the slightly larger emulation errors for $k$-modes beyond $0.2\,\hinvMpc$. The largest shift (with a mean value of $\sim 0.25\,\sigma_{X,\mathrm{true}}$) occurs for $c_4$ in case of the VDG model at $z= 0.9$. As we will show in Sec.~\ref{sec:reconstruction}, this is because the VDG model is more heavily affected by inaccuracies from the reconstruction of the anisotropic power spectrum and these inaccuracies are most notable for the small-scale hexadecapole. Since $c_4$ is mainly constrained by the hexadecapole, it absorbs these mismatches, resulting in the larger shifts. Finally, we note that the standard deviations for the distributions of the shifts are a little smaller than the value of the maximum difference (in units of $\sigma$) for $68\,\%$ of the samples found in Sec.~\ref{sec:relative_inaccuracies}, but they are generally consistent, suggesting that the latter can serve as a good indicator for the performance of the emulator in explorations of the likelihood.

It is instructive to check whether the largest inaccuracies from the emulator occur dominantly in certain parts of the cosmological parameter space. For that reason in Figs.~\ref{fig:EFT_diff_means_z0p9_b1} and \ref{fig:VIR_diff_means_z0p9_b1} we plot the shifts in the linear bias parameter for the EFT and VDG models at $z=0.9$, respectively, using all two-dimensional projections of the validation parameter set. We see indeed that for certain parameter combinations larger shifts (lighter colours) do not appear randomly, but in well separated regions: for instance, for the EFT model the most obvious separation occurs in the $\omega_b$ - $\omega_c$ parameter plane, showing that there are larger inaccuracies for large values of $\omega_b$ and simultaneously small values of $\omega_c$. A similar trend can also be observed for the VDG model, in which case there is an additional tendency for larger shifts in the $\omega_b$ - $A_s$ parameter plane. We obtain similar results for the other redshifts, emphasising that in central regions of the parameter space, in particular for values of $\omega_b$ close to the \emph{Planck} or Big Bang Nucleosynthesis priors, where one would preferentially sample the cosmological likelihood, the emulator performs best.

\subsubsection{Impact on credible regions}
\label{sec:credible_regions}

Finally, let us consider how well we can recover the 1-$\sigma$ credible regions for the parameters varied in the chains. This is more difficult to quantify precisely because, unlike the posterior means, the credible regions are more heavily affected by sampling noise, i.e., they carry a stronger dependence on the initial seed used for the MCMC at fixed convergence criterion. Ideally, we would therefore first quantify the sampling noise for each point in the validation set by running multiple chains with different initial seeds and constructing probability distributions of the 1-$\sigma$ credible region for each parameter, which could be compared between the exact and emulated models. However, as that procedure is very costly due to the need of running many individual chains, we settle on an approximate comparison only.

First we assume that the sampling noise is independent of cosmology and that the values of the 1-$\sigma$ confidence limits for each bias parameter are drawn from Gaussian distributions with means $\overline{\sigma}_{X,\mathrm{true/emu}}(\boldsymbol{\Theta})$ and variance $\sigma_{X, \mathrm{sampling}}^2$, where $\boldsymbol{\Theta}$ denotes the dependence on cosmological parameters contained in the validation set. The difference in the estimated 1-$\sigma$ confidence limits between the true and emulated models is then also Gaussian distributed with mean $\overline{\sigma}_{X,\mathrm{emu}}(\boldsymbol{\Theta}) - \overline{\sigma}_{X,\mathrm{true}}(\boldsymbol{\Theta}) = \Delta\overline{\sigma}_{X}(\boldsymbol{\Theta})$ and variance $2\sigma^2_{X,\mathrm{sampling}}$. In the case that $\Delta \overline{\sigma}_X(\boldsymbol{\Theta})$ is small or has negligible dependence on cosmology, we can regard each value $\Delta \sigma_X$ obtained at a different validation cosmology as being independently drawn from the same sampling noise distribution and we can interpret the offset of that distribution from zero mean as the accuracy with which we can recover the 1-$\sigma$ credible regions. On the other hand, if $\Delta \overline{\sigma}_X(\boldsymbol{\Theta})$ strongly depends on cosmology, the distribution of $\Delta \sigma_X$ values over the validation set will be broader and/or have a different shape, and so we cannot immediately assess the significance of the differences in $\sigma_X$ from the data we have generated.

In Fig.~\ref{fig:emu_comparison_stds} we plot the distributions of $\Delta \sigma_X$, normalised by the average over the entire validation set, $\left<\sigma_{X,\mathrm{true}}\right>$, for the two RSD models and the four different redshifts. In order to quantify the sampling noise we pick a $\Lambda$CDM cosmology with parameter values corresponding to the \emph{Planck} 2018 TT,TE,EE+lowE+lensing constraints, for which we run 1000 chains with the exact predictions for the EFT and VDG models at each redshift, and varying the same seven bias parameters as before, but with different initial seeds. We then determine $\sigma^2_{X,\mathrm{sampling}}$ in each case by fitting a Gaussian to the resulting distributions of $\sigma_{X}/\left<\sigma_X\right> - 1$, where $\left<\sigma_X\right>$ is the average of $\sigma_X$ over the 1000 different chains. This allows us to plot the reference sampling noise distributions (grey dashed lines in Fig.~\ref{fig:emu_comparison_stds}) as Gaussians with variance $2\sigma_{X,\mathrm{sampling}}^2$ and mean given by the median of the EFT distributions (blue histograms).

From these plots we observe that many of the histograms over the validation set are indeed consistent with sampling noise and moreover, in those cases the differences between $\sigma_{X,\mathrm{true}}$ and $\sigma_{X,\mathrm{emu}}$ are at the per-cent level, which is insignificant compared to the spread due to sampling noise. Some parameters, in particular $b_1$, $N_0$, and $c_0$, display broader or skewed distributions, suggesting that in those cases the cosmology dependence of the differences of $\sigma_{X}$ is stronger. In those cases we can only deduce that on average the differences are still at the per-cent level (see median values), but it is not possible to judge their significance. The deviations from the sampling noise distributions are largest at $z=0.9$, where the synthethic measurements uncertainties are smallest and hence where the emulation inaccuracies carry the strongest weight, while going to $z=1.8$ gives again very good consistency with sampling noise for all parameters. We do not find any significant differences between the two RSD models.

\subsection{Results for analysis with varying cosmological parameters}
\label{sec:fixed_cosmo_validation}

\begin{figure*}
  \centering
  \includegraphics{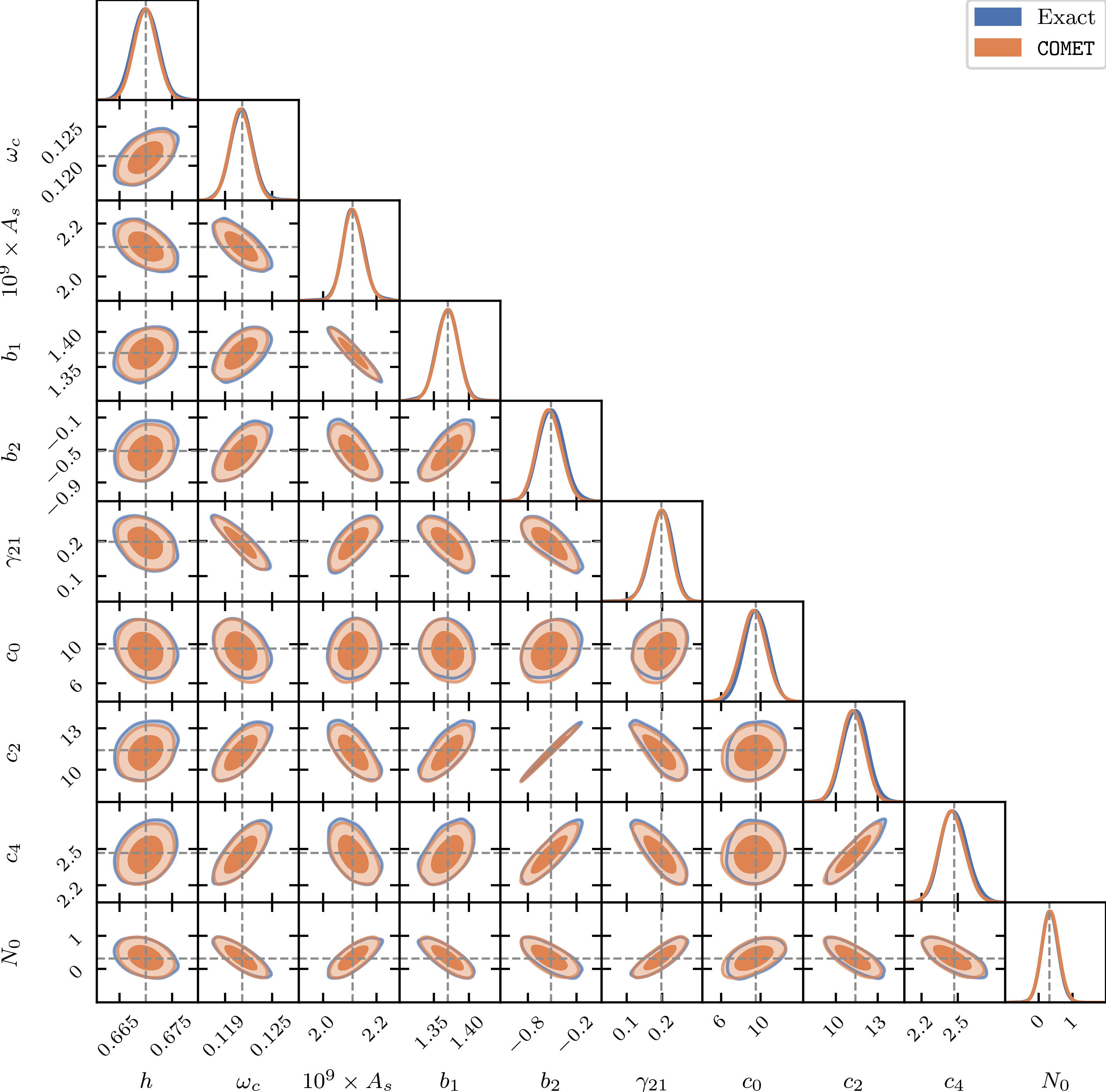}
  \caption{Comparison of the 2d and 1d marginalised posteriors obtained from running MCMC with the exact model and \emu\ for the EFT. The results shown stem from the synthetic data set at redshift $z=0.9$, as described in Sec.~\ref{sec:fixed_cosmology} and using a volume corresponding to 10 times the volume of a \emph{Euclid} redshift shell. The value of $\gamma_2$ was constrained to the excursion set relation of \citet{EggScoCro2011}, while $c_{\rm nlo}$ was fixed to the fiducial value (see Table~\ref{tab:validation_params}). The three counterterm parameters $c_0$, $c_2$ and $c_4$ are given in units of $(\hMpc)^2$. Vertical and horizontal dashed lines indicate the fiducial parameter values.}
  \label{fig:emu_comparison_all_params_EFT_z0p9}
\end{figure*}

Finally, we analyse the synthetic power spectrum multipoles described in Sec.~\ref{sec:fixed_cosmology}. As before, we run two chains, one with the exact model, the other using \emu, but instead of keeping the cosmological parameters fixed as in the previous section, we now also vary $h$, $\omega_c$ and $A_s$, setting only $n_s$ and $\omega_b$ to their fiducial values. Out of the full set of bias parameters we include $b_1$, $b_2$, $\gamma_{21}$, $c_0$, $c_2$, $c_4$, and $N_0$ in the chains, fixing $\gamma_2$ and $c_{\rm nlo}$ to the values used in the generation of the synthetic data. Since explorations of the likelihood with varying cosmological parameters is much more computationally expensive for our exact model code, we limits ourselves here to only a single case per redshift.

The chains are run using \texttt{MultiNest} \citep{FerHob0802,FerHobBri0910,FerHobCam1911} with 1800 live points in case of \emu\ and a standard Metropolis-Hastings sampler with a total number of 42,000 accepted steps in case of the exact model. After processing these chains with \texttt{getdist} \citep{Lew1910} we obtain the 2d marginalised posteriors for the full set of parameters at redshift $z = 0.9$ shown in Fig.~\ref{fig:emu_comparison_all_params_EFT_z0p9}, where the results based on the exact model correspond to the blue contours, the ones based on \emu\ to the orange contours. We see that the agreement between the posteriors of the two models is close to perfect: the mean posterior values of all parameters are almost identical and any occuring shifts are well below the 1-$\sigma$ level, while the 1-$\sigma$ and 2-$\sigma$ credible regions are equally well recovered. Some slight differences can be observed in the tails of the posterior distributions, but these are most sensitive to the sampling routines and therefore most likely caused by differences in the two samplers used here. Although not shown, we find qualitatively very similar results at the three remaining redshifts, so that these findings confirm our results from Sec.~\ref{sec:LCDM_validation} at fixed cosmology. 

\subsection{Reconstruction of anisotropic power spectrum}
\label{sec:reconstruction}

In this section we report the impact of inaccuracies caused by reconstructing the full anisotropic power spectrum from the monopole, quadrupole and hexadecapole only, as well as the approximate inclusion of the $\ell=6$ multipole discussed in Sec.~\ref{sec:emu_reconstruction}. To that end we make again use of the validation set described in Sec.~\ref{sec:LCDM_validation_set} and compute the first three multipoles for each validation cosmology using the exact model (i.e., without any input from the emulator), but without inclusion of Alcock-Paczynski distortions and, in case of the VDG model, without the FoG damping term. Like for our emulator, we then reconstruct the anisotropic power spectrum from those multipole moments, apply Alcock-Pacyznski distortions and FoG damping, and as a final step evaluate the observed multipoles. We can compare these predictions with those that do not make use of the multipole reconstruction in order to determine the differences as a function of the cosmological parameters.

\begin{figure*}
  \centering
  \includegraphics{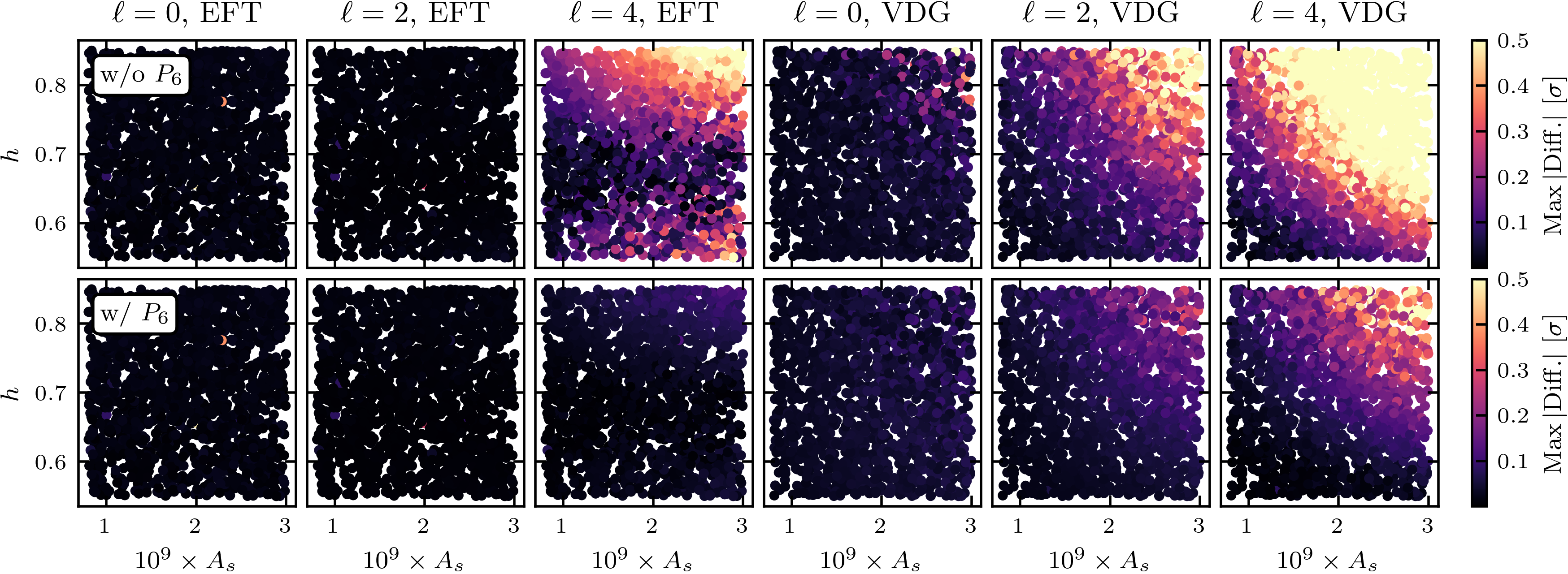}
  \caption{Maximum differences up to $k_{\mathrm{max}} = 0.3\,\hinvMpc$ (in units of the standard deviation of our synthetic data set, see Sec.~\ref{sec:synthetic_data}) for each point of the validation set between the exact computation of the power spectrum multipoles and when the anisotropic power spectrum before application of Alcock-Paczynski distortions and FoG damping is obtained from a Legendre decomposition truncated at finite order. In the top row the latter is based on the first three non-zero multipoles, while in the bottom row we include the $\ell=6$ multipole moment evaluated at fixed shape parameters (see Sec.~\ref{sec:emu_reconstruction} for details). Both sets of predictions are generated without any input from the emulator.}
  \label{fig:recon_inacc_P6}
\end{figure*}

This is shown in the top row of Fig.~\ref{fig:recon_inacc_P6}, where we plot the maximum difference (taken over all scales up to $k_{\mathrm{max}} = 0.3\,\hinvMpc$) in units of the synthetic standard deviations for each validation point at $z=0.9$, projected into the $h$ - $A_s$ parameter plane. The first three panels depict these differences for the monopole to hexadecapole for EFT model, the next three for the VDG model. We note that for the EFT model there is virtually no impact on the monopole and quadrupole, even for measurement uncertainties corresponding to 10 times the volume contained in a \Euclid\ redshift shell. The situation is different for the hexadecapole, where we obtain maximum differences of up to $\sim 0.5\sigma$, in particular for large and small values of $h$. On the other hand, for the VDG model the effect is noticeably larger with maximum differences going well beyond $0.5\sigma$ for the hexadecapole (up to $\sim 2.5\sigma$) and also more significant inaccuracies for the monopole and quadrupole, preferentially for large values of $h$ and $A_s$. This happens because the FoG damping factor carries a significant additional line-of-sight dependence, which amplifies the contributions from higher multipole moments not included in the reconstruction.

The lower panels of Fig.~\ref{fig:recon_inacc_P6} shows how the maximum difference improve when the $\ell=6$ multipole, evaluated at fixed shape parameters corresponding to the \emph{Planck} 2018 TT,TE,EE+lowE+lensing values, is included in the Legendre expansion. We see that even when using this approximation the inaccuracies are significantly reduced for both RSD models. Specifically, they now stay below $\sim 0.2\sigma$ for the hexadecapole of the EFT model, while only about $5\,\%$ of the validation samples reach maximum difference larger than $\sim 0.2\sigma$ and $0.35\sigma$ for the quadrupole and hexadecapole of the VDG model, respectively. Returning to the larger shifts in the counterterm parameter $c_4$ that we noticed in Sec.~\ref{sec:shifts_posterior_means} for the VDG model, we find that they depend on cosmology in a very similar way as the differences in the bottom right panel of Fig.~\ref{fig:recon_inacc_P6}, implying that these shifts are caused by the remaining reconstruction inaccuracies in the hexadecapole. However, we stress that these are not only negligible, they also occur in a region of the cosmology parameter space that is not relevant for likelihood explorations.

%%%%%%%%%%%%%%%%%%%%%%%%%%%%%%%%%%%%%%%%%%%%%%%%%%%%%%%%%%%%%%%%%%%%%%%%%%%%%%%%%%%%%%%%%%%%%%%%%%%%%%%%%%%%%%%%%%%%%%%%%%%%%%%%%%%%%%%%%%%%%%%%%%%%%%
\section{Conclusions}
\label{sec:conclusions}

\subsection{Summary}

We have presented \emu, an emulator of the galaxy power spectrum multipoles in redshift-space based on two different perturbation theory models: the EFT model as employed in the analyses of \citet{IvaSimZal2005,dAmGleKok2005}, which fully expands the real- to redshift-space mapping, and the VDG model, which models the impact of small-scale velocity differences via a non-perturbative damping function \citep{Sco0410,TarNisSai1009,SanScoCro1701}. The leading idea that was driving the design of our emulator was to minimise the emulation parameter space, in order to reach an optimal compromise between computation time and accuracy. For that reason we have adopted the evolution mapping approach of \citet{SanRuiGon2108} and trained \emu\ internally in units of $\Mpc$ over the range $0.0007\,\invMpc$ to $0.35\,\invMpc$ using only the shape parameters $\omega_b$, $\omega_c$, $n_s$, in addition to $\sigma_{12}$ and the growth rate $f$. In this way we are able to support a broad set of evolution parameters, specifically $h$, $A_s$, $\Omega_K$, $w_0$, and $w_a$, by mapping them to the corresponding values of $\sigma_{12}$ and $f$ at any given redshift (up to an upper limit of $z \sim 3$, imposed by our chosen range for $\sigma_{12}$). Furthermore, we emulate all independent contributions that arise from the galaxy bias expansion separately, which precludes the associated parameters from the emulator parameter space, and apply AP distortions and the effective damping function in case of the VDG model in a separate step. This gives \emu\ the flexibility to support any fiducial background cosmologies and arbitrary functional forms of the damping term. A single evaluation of the monopole, quadrupole and hexadecapole at ${\cal O}(100)$ scales takes about $10\,\mathrm{ms}$ when executed on a single CPU.

Using a series of validation tests, we verified that \emu\ does not introduce any relevant loss in accuracy in comparison to the exact perturbation theory models. We constructed a large validation set consisting of 1500 synthetic power spectrum measurements at four different redshifts between $z=0.9$ and $z=1.8$, covering the five-dimensional cosmological parameter space: $\omega_b$, $\omega_c$, $n_s$, $h$ and $A_s$. Adopting a fixed set of galaxy bias parameters at each redshift and in case of the EFT model, we find that the relative inaccuracies stay below $0.08(0.14)\,\%$, $0.11(0.17)\,\%$ and $0.30(1.33)\,\%$ for $68(95)\,\%$ of the validation samples and the monopole, quadrupole, and hexadecapole, respectively, up to $k_{\rm max} = 0.3\,\hinvMpc$. We further generated statistical uncertainties for our synthetic measurements using Gaussian covariances and specifics tied to the \emph{Euclid} survey, but with a tenfold increase in volume. In units of these uncertainties the emulation errors are below $0.24(0.42)\,\sigma$, $0.10(0.17)\,\sigma$, and $0.05(0.10)\,\sigma$. We then ran MCMC, varying a set of seven bias parameters both for \emu\ and the exact model and analysing all three synthetic multipoles for each validation sample, finding that the shifts in the mean posterior values closely match the emulation inaccuracies in units of $\sigma$. Additionally, we find that the $68\,\%$ credible regions are very well recovered and any occurring differences are typically smaller than the MCMC sampling noise. For the VDG model the performance is very similar, apart from a slightly less accurate hexadecapole (see discussion in Sec.~\ref{sec:reconstruction}), but without negative effects on the recovery of the posteriors. Finally, we constructed one more synthetic collection of measurements (again at the same four refshifts) for a fixed set of cosmological and bias parameters, which we used to demonstrate that even when running chains over the full parameter space (including cosmological parameters) there is no appreciable difference in the resulting posterior distributions between the exact model and \emu.

While we have explicitly demonstrated that the emulator is accurate up to scales of at least $k_{\rm max} = 0.3\,\hinvMpc$, we caution that this does not have to apply to the underlying theoretical model itself. The validity of the one-loop perturbation theory depends on the relative size of the neglected non-linear corrections, as well as the amplitude of the galaxy bias parameters, and is thus a function of redshift and the particular galaxy sample under consideration. Any application to real data must therefore be preceded by a thorough study of the model's robustness under changes of the maximum scale cuts --- a task that is ideally suited for \emu\ due to its superior computational efficiency.

We leave two important extensions of \emu\ to forthcoming publications: Firstly, the power spectrum models for both the EFT and VDG can be easily transformed to configuration space and thus, using the same emulation technniques as presented here, we would also be able to make fast predictions of the two-point correlation function multipoles. Secondly, the new generation of galaxy surveys is expected to improve our current constraints on the masses of neutrinos, which is why it is particularly important to be able to predict galaxy clustering statistics as a function of non-zero neutrino masses, unlike we have assumed here.

\subsection{Comparison to related emulators in the literature}

While most galaxy clustering emulators that have been presented in the literature so far have been built from simulations and focus on the non-linear regime, there are two sets of works, which are closely related to what we have presented here and which we want to briefly compare against. In particular, \citet{DonKoyBeu2202,DeRCheWhi2204} have presented two perturbation theory emulators of the galaxy power spectrum multipoles. The former, \texttt{EFTEMU}, is based on the \texttt{PyBird} code \citep{dAmGleKok2005}, which implements perturbation theory expressions identical to those described in Sec.~\ref{sec:models} for the EFT model apart from slight differences in the infrared resummation procedure\footnote{A comparison between \texttt{CLASS-PT} \citep{ChuIvaPhi2009}, which uses the same infrared resummation technique as discussed in Sec.~\ref{sec:IR-resummation}, and \texttt{PyBird} has been presented in \citet{NisdAmIva2012}, finding very similar results.} and the definition of the galaxy bias parameters (see Sec.~\ref{sec:bias} for a conversion), while the latter, \texttt{EmulateLSS}, is based on the Lagrangian perturbation theory model of \citet{CheVlaCas2103}. Although sharing the same goal, there are a number of differences between these two emulators and \emu. On the one hand, this concerns the parameter space and the range of scales for which they provide predictions: \texttt{EFTEMU} supports the five cosmological parameters $\omega_b$, $\omega_c$, $h$, $A_s$, and $n_s$ (i.e., it does not allow for deviations from the equation of state of a cosmological constant or for non-flat cosmologies), and with the exception of $n_s$ all of these have smaller ranges than in \emu. \texttt{EmulateLSS} is even more restrictive as it also fixes the spectral index and does not cover the full galaxy bias and counterterm parameter space, ignoring for instance the second- and third-order tidal bias parameters. Moreover, both of these emulators make predictions for a fixed fiducial background cosmology and at fixed redshifts (each new fiducial background cosmology or redshift would require generating a new set of training data) and while \texttt{EmulateLSS} gives predictions in the range of scales from $0.001\,\hinvMpc$ to $0.5\,\hinvMpc$, \texttt{EFTEMU} has a more limiting range with a maximum wavemode of $0.19\,\hinvMpc$. On the other hand, the prediction accuracies for the three multipoles quoted by \citet{DeRCheWhi2204} are only slightly worse than what we have determined for \emu, and also \citet{DonKoyBeu2202} find a sub-percent accuracy for fixed BOSS-like galaxy bias parameters, making it comparable with \emu in this regard (albeit on a more limited range of scales). Finally, both \texttt{EFTEMU} and \texttt{EmulateLSS} use neural networks instead of Gaussian processes for the emulation procedure and their computation time therefore does not depend on the size of the training set. This leads to a better computational performance than \emu\ by about one order of magnitude based on the fact that both works quote a computation time of $\sim 1\,\mathrm{ms}$ (we stress, however, that eventually the computational bottleneck will lie elsewhere, e.g. in the computation of the $\chi^2$ or, for joint chains with the bispectrum, the computation of the bispectrum model).

The methodology discussed in \citet{ZenAngPel2101,AriAngZen2104,KokDeRChe2107} follows the same perturbative galaxy bias expansion as presented in Sec.~\ref{sec:bias} (although the third order term related to the parameter $\gamma_{21}$ is neglected) and they have presented emulators for the various contributions to the galaxy power spectrum based on either perturbation theory \citep{AriAngZen2104} like here, or on simulations in order to extend the predictions into the non-linear regime \citep{ZenAngPel2101,KokDeRChe2107}. All three works consider scales up to $1\,\hinvMpc$, but cover somewhat more restrictive redshift ranges than \emu, $z_{\rm max} = 1.5$ \citep{AriAngZen2104,ZenAngPel2101} and $z_{\rm max} = 2$ \citep{KokDeRChe2107}. They also emulate over somewhat different cosmological parameter spaces: the former two also cover an eight-dimensional cosmology parameter space, but instead of $\Omega_K$ account for the neutrino mass, while in the latter they use a seven-dimensional parameter space that includes the effective number of relativistic species, but does not allow for either $\Omega_K$ or a dynamical dark energy equation of state. Moreover, they generally quote an emulation accuracy of $\sim 1\,\%$ --- one order of magnitude larger than our findings here, but for this comparison one should bear in mind that these results carry a dependency on the nature of the validation tests. While these three works only make predictions for the galaxy power spectrum in real-space, \citet{PelStuAng2208} extends their formalism in order to account for redshift-space distortions. They show that using a halo displacement field extracted from simulations and a phenomenological finger-of-god model accounting for the velocity dispersion of satellite galaxies and satellite fraction, it is possible to recover power spectrum multipoles up to $k \sim 0.6\,\hinvMpc$ that are accurate within measurement uncertainties corresponding to a volume of $3 (h^{-1}\,\mathrm{Gpc})^3$.

\subsection{Functionality of the \emu\ package}

\emu\ is a freely available \texttt{Python} package (\url{https://gitlab.com/aegge/comet-emu}), which can be installed via \texttt{pip} and all required tables and emulators are downloaded automatically (re-training the emulators is not necessary). We currently provide emulators for the real-space power spectrum and the redshift-space power spectrum multipoles for the EFT model\footnote{The VDG model will be released as part of an upcoming publication.}. Predictions can be made using either the $\Mpc$ or $\hMpc$ unit systems and for either the native parameter space or for a given dark energy model. The former consists of the three shape parameters $\omega_b$, $\omega_c$ and $n_s$, in combination with $\sigma_{12}$ and $f$, while in case of the latter one needs to specify the evolution parameters $h$, $A_s$, $\Omega_K$, $w_0$ and $w_a$ at some redshift $z$ instead of $\sigma_{12}$ and $f$. \emu\ accepts an arbitrary range of scales, but for those scales outside of the range for which we trained the emulators ($k \in [0.0007,\,0.35]\,\invMpc$), we apply a power law extrapolation. Further features of the package include:
\begin{itemize}
\item the prediction of the linear matter power spectrum, with or without application of infrared resummation (BAO damping)
\item the prediction of the real-space tree-level galaxy bispectrum and the tree-level galaxy bispectrum multipoles in redshift-space
\item the computation of the Gaussian covariance matrices of the power spectrum and bispectrum multipoles (both, real- and redshift-space)
\item the computation of the $\chi^2$ for a given set of measurements and their covariance matrix for arbitrary $k_{\rm max}$ scale cuts; we also provide a functionality that drastically speeds up the $\chi^2$ computation in case of fixed cosmological parameters (e.g., when running a chain over only galaxy bias parameters)
\item the possibility to choose between different galaxy bias bases (see Sec.~\ref{sec:bias})
\item exact treatment of discreteness and finite bin width effects for the power spectrum multipoles
\end{itemize}
More information and some tutorials can be found on the documentation pages: \url{https://comet-emu.readthedocs.io/en/latest/index.html}.

%%%%%%%%%%%%%%%%%%%%%%%%%%%%%%%%%%%%%%%%%%%%%%%%%%%%%%%%%%%%%%%%%%%%%%%%%%%%%%%%%%%%%%%%%%%%%%%%%%%%%%%%%%%%%%%%%%%%%%%%%%%%%%%%%%%%%%%%%%%%%%%%%%%%%%
\section*{Acknowledgements}

We thank the anonymous referee for useful comments, in particular the suggestion that led to the inclusion of Appendix \ref{sec:discreteness}. AE is supported at the Argelander Institut f\"ur Astronomie by an Argelander Fellowship. BCQ and MC acknowledge support from the Spanish Ministerio de Ciencia e Innovación under grant PGC2018-102021-B-I00, and BCQ additionally acknowledges support from a PhD scholarship from the Secretaria d’Universitats i Recerca de la Generalitat de Catalunya i del Fons Social Europeu. AGS acknowledges the support of the Excellence Cluster ORIGINS, which is funded by the Deutsche Forschungsgemeinschaft (DFG, German Research Foundation) under Germany's Excellence Strategy - EXC-2094 - 390783311. This research made use of matplotlib, a Python library for publication quality graphics \citep{Hun0705}.

%%%%%%%%%%%%%%%%%%%%%%%%%%%%%%%%%%%%%%%%%%%%%%%%%%%%%%%%%%%%%%%%%%%%%%%%%%%%%%%%%%%%%%%%%%%%%%%%%%%%%%%%%%%%%%%%%%%%%%%%%%%%%%%%%%%%%%%%%%%%%%%%%%%%%%
\section*{Data Availability}

The data underlying this article will be shared on reasonable request to the corresponding author.



\bibliographystyle{mnras}
\bibliography{references} % if your bibtex file is called example.bib

%\begin{thebibliography}{99}
%\end{thebibliography}

%%%%%%%%%%%%%%%%%%%%%%%%%%%%%%%%%%%%%%%%%%%%%%%%%%

%%%%%%%%%%%%%%%%% APPENDICES %%%%%%%%%%%%%%%%%%%%%

\appendix

\section{Perturbation theory kernels}
\label{sec:loop_integrals_expressions}

In the formulation used by \emu, the perturbative term of both redshift-space models, EFT and VDG, can be written in the following way,
\begin{equation}
    \begin{split}
      P_{gg}(k,\mu) = \;& P^{\rm tree}_{gg,\rm SPT}(k,\mu) +   P^{\rm 1-loop}_{gg,\rm SPT}(k,\mu) \\
                        & + P_{gg}^{\rm{stoch}}(k,\mu)  +  P_{gg}^{\rm{ctr}}(k,\mu)\,,
    \end{split}
  \end{equation}
where the individual contribution are given by
\begin{equation}
    P_{gg,\rm SPT}^{{\rm{tree}}}(k,\mu) = Z_1(\kv)^2 \, \Plin(k),  
\end{equation}
\begin{equation}
    \begin{split}
        P_{gg,\rm SPT}^{{\rm{1-loop}}}(k,\mu) = \;& P_{gg,22}(k,\mu) + P_{gg,13}(k,\mu) = \\
        = \;& 2\int_{\qv} Z_2(\boldsymbol{q},\kv-\qv)^2\,\Plin(|\kv-\qv|)\,\Plin(q)  \\
        & + 6\,Z_1(\kv)\,\Plin(k)\int_{\qv}Z_3(\qv,-\qv,\kv)\,\Plin(q)\,,
    \end{split}
    \label{eq:Pgg_1loop}
\end{equation}
\begin{equation}
  P_{\rm{gg}}^{{\rm{stoch}}}(k,\mu) = \frac{1}{\bar{n}}\left\{N_0^P+k^2\left[N_{20}^P+N_{22}^P\mathcal{L}_2(\mu)\right]\right\}\,,
\end{equation}
\begin{equation}
    \begin{split}
        P_{gg}^{{\rm{ctr}}}(k,\mu) = \;& P_{gg}^{{\rm{ctr,LO}}}(k,\mu) + P_{gg}^{{\rm{ctr,NLO}}}(k,\mu) = \\
        = \; & -2\left[c_0+c_2\mathcal{L}_2(\mu)+c_4\mathcal{L}_4(\mu)\right]k^2\Plin(k) \\
        & + c_\mathrm{nlo}f^4\mu^4\,\,k^4\,Z_1(\kv)^2\,\Plin(k)\,.
    \end{split}
\end{equation}
Here, the redshift-space kernels $Z_n$ are defined as
\begin{equation}
    Z_1(\kv) = b_1+f\mu^2,
\end{equation}
\begin{equation}
    \begin{split}
        Z_2(\kv_1,\kv_2) = \;& {\cal K}_2(\kv_1,\kv_2) + f\mu^2 G_2(\kv_1,\kv_2) \,+ \\
        & + \frac{1}{2}fk\mu \left[\frac{\mu_1}{k_1}\left(b_1+f\mu_2^{\,2}\right)+\frac{\mu_2}{k_2}\left(b_1+f\mu_1^{\,2}\right)\right]\,,
    \end{split}
\end{equation}
\begin{equation}
    \begin{split}
        Z_3(\kv_1,\kv_2,\kv_3) = \;& {\cal K}_3(\kv_1,\kv_2,\kv_3) + f\mu^2 G_3(\kv_1,\kv_2,\kv_3)  \\
        & + \frac{1}{2}f^2k^2\mu^2\frac{\mu_2\,\mu_3}{k_2\,k_3}\left(b_1+f\mu_1^{\,2}\right) \\
        & + fk\mu\frac{\mu_3}{k_3}\left[b_1F_2(\kv_1,\kv_2)+f\mu_{12}^{\,2}G_2(\kv_1,\kv_2)\right]  \\
        & + fk\mu\frac{\mu_{23}}{k_{23}}\left(b_1+f\mu_1^{\,2}\right)G_2(\kv_2,\kv_3)  \\
        & +fk\mu\frac{\mu_1}{k_1}\left[\frac{b_2}{2}+\gamma_2K(\kv_2,\kv_3)\right]\,,
    \end{split}
\end{equation}
where the $Z_3(\kv_1,\kv_2,\kv_3)$ needs to be symmetrised over its arguments. The real-space galaxy kernels ${\cal K}_n$ read
\begin{equation}
  \label{eq:K2}
    {\cal K}_2(\kv_1,\kv_2)=b_1F_2(\kv_1,\kv_2)+\frac{b_2}{2}+\gamma_2\, K(\kv_1,\kv_2)\,,
\end{equation}
\begin{equation}
  \label{eq:K3}
    \begin{split}
        {\cal K}_3(\kv_1,\kv_2,\kv_3)= \;& b_1F_3(\kv_1,\kv_2,\kv_3) + b_2F_2(\kv_1,\kv_2) \\
        &\hspace{-5em}+2\gamma_2K(\kv_1,\kv_2+\kv_3)G_2(\kv_2,\kv_3) \\
        &\hspace{-5em}+2\gamma_{21}K(\kv_1,\kv_2+\kv_3)\,K(\kv_2,\kv_3)\,,
    \end{split}
\end{equation}
with 
\begin{equation}
    K(\kv_1,\kv_2)=\frac{\left(\kv_1\cdot\kv_2\right)^2}{k_1^{\,2}k_2^{\,2}}-1\,.
 \end{equation}
 The kernel ${\cal K}_3(\kv_1,\kv_2,\kv_3)$ also needs to be symmetrised over its arguments.

\section{Binning and discreteness effects in estimates of the power spectrum}
\label{sec:discreteness}

The power spectrum multipoles are estimated from density grids in Fourier space by taking the square of the Fourier coefficients at each wavevector $\bk$ and multiplying by the respective Legendre polynomial \citep[for more detail, see][]{Sco1510}. These estimates are then summarised into measurements at multiple wavemode bins, $k_i$, by averaging over all wavevectors that fall into a spherical shell centred on $k_i$ with some specified bin width $\Delta k$. Both, the finite bin width as well as the discreteness of the Fourier grid, introduce differences compared to the theoretical power spectrum multipoles as defined in Sec.~\ref{sec:emulator}, which are evaluated at fixed $k$ and by integrating over continuous values of $\mu$, the orientation of $\bk$ with respect to the line-of-sight (see e.g. Eq.~\ref{eq:emu.Pell}). This can be accounted for by averaging the theoretical power spectrum $P(k,\mu)$ in the same way as done for the measurements \citep[e.g.][]{TarNisBer1304}. The ``observed'' theoretical multipoles are thus given by
\begin{equation}
  \label{eq:Pell_obs}
  P_{\ell}^{\rm obs}(k_i) = \frac{2\ell + 1}{N_k} \sum_{|\bk| \in [k_i - \Delta k/2, k_i + \Delta k/2]} P(k,\mu)\,{\cal L}_{\ell}(\mu)\,,
\end{equation}
where $N_k$ are the total number of wavevectors per spherical shell.

\begin{figure}
  \centering
  \includegraphics[width=0.98\columnwidth]{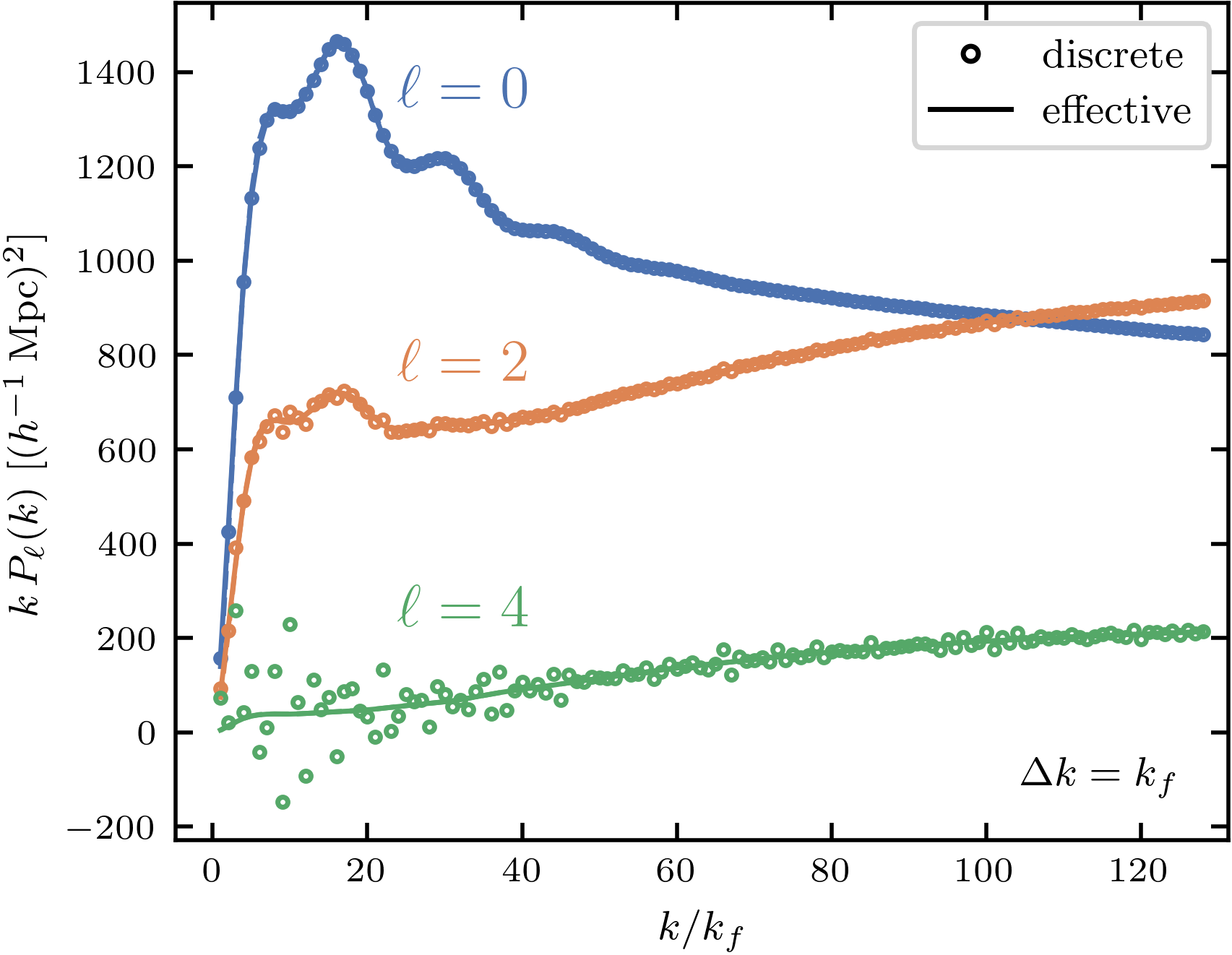}
  \caption{Comparison of the power spectrum multipole predictions evaluated at the effective wavemodes (lines) and averaged over discrete set of $k$ and $\mu$ values Eq.~(\ref{eq:Pell_obs}) (circles). The bin width was assumed to be $k_f$, with $k_f \approx 0.0042 \hinvMpc$.}
  \label{fig:Pell_discrete_effective}
\end{figure}

Instead of averaging the full power spectrum, a common approximation is to evaluate the power spectrum multipoles at the \emph{effective} wavemodes, defined as
\begin{equation}
  \label{eq:keff}
  k_{i,\rm eff} \equiv \frac{1}{N_k} \sum_{|\bk| \in [k_i - \Delta k/2, k_i + \Delta k/2]} |\bk|\,,
\end{equation}
which can partially account for the finite bin width. In Fig.~\ref{fig:Pell_discrete_effective} we compare Eq.~(\ref{eq:Pell_obs}) (circles) against the power spectrum multipoles evaluated at the effective wavemodes (lines) for a Fourier grid with fundamental frequency, $k_f = 2\pi/L$, corresponding to a box size of $L = 1500\,\hMpc$ and bin width $\Delta k = k_f$. While the discreteness effect is barely noticeable for the monopole, it already gives rise to per-cent level differences in the quadrupole, and even more significant effects in the hexadecapole. This is particularly the case for bins which are close to the fundamental frequency, where there are consequently only a small number of wavevectors per spherical shell, but for the hexadecapole differences up to $\sim 10\,\%$ persist up to $k \sim 100\,k_f$.

We have implemented the exact bin average for arbitrary bin widths (with linear spacing) in \emu\ by computing Eq.~(\ref{eq:Pell_obs}) over the anisotropic power spectrum reconstructed from the emulated quantities (see Sec.~\ref{sec:emu_reconstruction}). We have verified that the reconstruction of the anisotropic power spectrum from a finite number of multipoles does not introduce any appreciable inaccuracies when performing the bin average, as was also the case without bin average (see Sec.~\ref{sec:reconstruction}). In order to speed up the computation of the binned predictions, we first find all discrete values of $k$ and $\mu$ for each bin and then perform the following rounding operations
\begin{align}
  k &\approx \left\lfloor 10\frac{k}{\Delta k}\right\rceil\,\frac{\Delta k}{10}\,, \\
  \mu &\approx 10^{-3}\,\left\lfloor 10^3\mu\right\rceil\,,
\end{align}
where $\lfloor x \rceil$ denotes the nearest integer. Finally, we determine all unique combinations of $k$ and $\mu$, as well as the number of times they appear, so that we can evaluate Eq.~(\ref{eq:Pell_obs}) only by averaging over those unique combinations, using their respective number of occurances as weights. While this approximation only leads to negligible inaccuracies, it reduces the computation time and means that in the above example we can evaluate the three discretely averaged multipoles up to $k \sim 60\,k_f$ without any additional computational cost compared to the standard evaluation. That means the discrete average can be straightforwardly included also in any likelihood analysis.

\section{Statistics of emulation inaccuracies at various redshifts}
\label{sec:app_cumulative_histograms}

\begin{figure*}
  \centering
  \includegraphics{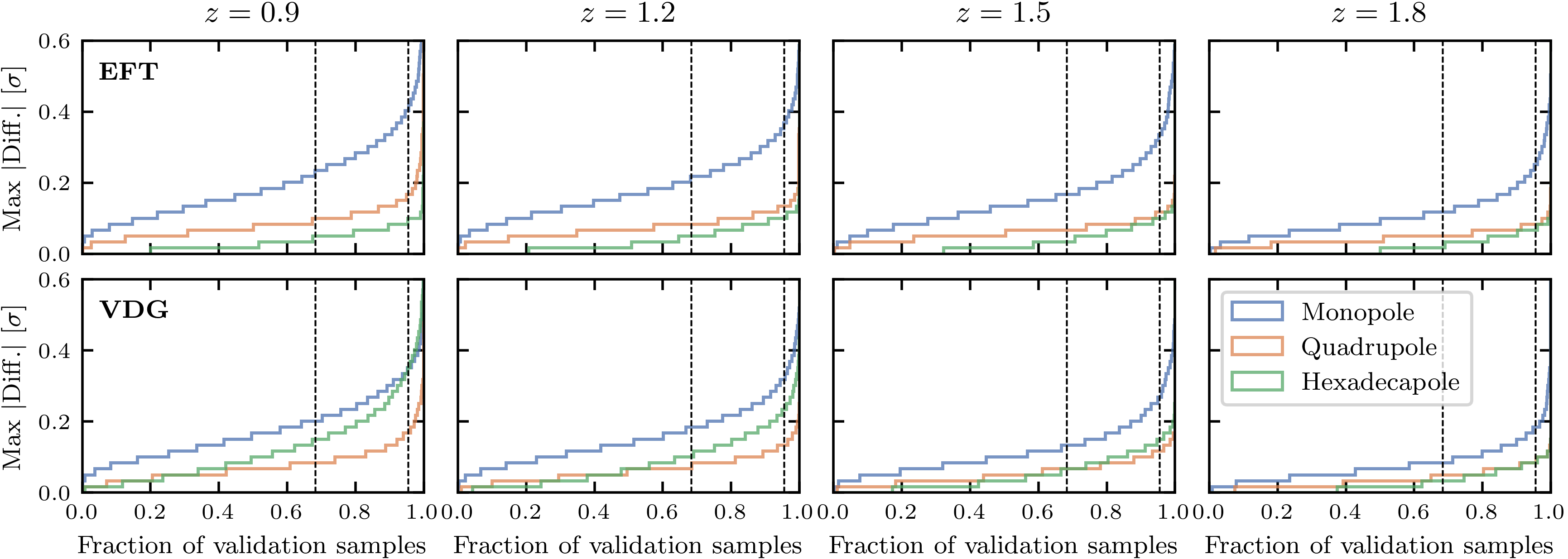}
  \caption{Cumulative histograms (over the full validation set) of the maximum absolute difference between the power spectrum multipoles computed in the exact model and with our emulator over a range of scales from $0.001\,\hinvMpc$ and $0.3\,\hinvMpc$. Differences are shown in units of the standard deviation of the synthetic data set described in Sec.~\ref{sec:LCDM_validation_set}. Each column corresponds to a different redshift, while the upper panels show results for the EFT model, lower panels for the VDG model. The dashed horizontal lines indicate $68\,\%$ and $95\,\%$ of the validation samples.}
  \label{fig:emu_comparison_max_diff_hist}
\end{figure*}

\begin{table*}
  \centering
  \caption{Fiducial values of number densities, galaxy bias and counterterm parameters, used in the generation of the synthetic data sets described in Sec.~\ref{sec:fixed_cosmology}.}
  \begin{tabular}{cccccccccc}
    \hline
    $z$ & $10^3 \times \bar{n}$ [$(\hMpc)^{-3}$] & $b_1$ & $b_2$ & $\gamma_{21}$ & $c_0$ [$(\hMpc)^2$] & $c_2$ [$(\hMpc)^2$] & $c_4$ [$(\hMpc)^2$] & $c_{\rm nlo}$ [$(\hMpc)^4$] & $N^P_0$ \\ \hline
    0.9 & 2.043 & 1.370 & -0.514 & 0.197 & 9.542 & 11.390 & 2.469 & 12.972 & 0.315 \\
    1.2 & 1.029 & 1.734 & -0.193 & 0.354 & 15.521 & 17.888 & 4.057 & 78.367 & 0.046 \\
    1.5 & 0.585 & 2.024 & 0.443 & 0.239 & 11.937 & 18.168 & 3.745 & 33.842 & -0.042 \\
    1.8 & 0.313 & 2.476 & 0.563 & -0.112 & 15.377 & 13.322 & 2.958 & -64.369 & 0.296 \\
    \hline
  \end{tabular}
  \label{tab:validation_params}
\end{table*}

For the sake of completeness, in Fig.~\ref{fig:emu_comparison_max_diff_hist} we show the cumulative histograms of the maximum absolute differences between the emulator and exact model for all four redshifts of our synthetic data sets described in Sec.~\ref{sec:LCDM_validation_set}. As in Sec.~\ref{sec:relative_inaccuracies}, the maximum differences for each multipole are obtained over a range of scales between $0.001\,\hinvMpc$ and $0.3\,\hinvMpc$ and overall the plot demonstrates that we get qualitatively a very similar behaviour for the validation samples at higher redshifts as for the one at $z = 0.9$ that was presented in the main text. When expressing the maximum differences in terms of the standard deviation we have adopted for our synthetic measurements, we still find the most significant discrepancies in the monopole. The fact that these decrease, however, with increasing redshift is due to the larger relative errors, $\sigma_{\ell}(k)/P_{\ell}(k)$, at higher redshifts, which are caused by our assumption of decreasing number densities and the resulting bigger contributions from shot noise to $\sigma_{\ell}$. For the EFT model we moreover see that the smallest discrepancies always occur in the hexadecapole, while, as explained in Sec.~\ref{sec:relative_inaccuracies}, they can be significantly larger in case of the VDG model. At $z=0.9$ they exceed those of the quadrupole and with increasing redshift they become comparable. These findings are fully compatible with our results on the shifts of mean parameter values in Sec.~\ref{sec:shifts_posterior_means}, where we have seen that the spread in the shifts become smaller for the higher redshift validation samples. This is, in particular, true for the counterterm parameter $c_4$, which was most affected by the inaccuracies in the hexadecapole at redshift $z=0.9$.

\section{Parameter values for fixed validation set}
\label{sec:validation_params}

In Table~\ref{tab:validation_params} we collect the values of the galaxy bias and counterterm parameters that were used to generate a set of synthetic power spectrum multipole measurements from the EFT model at four different redshifts. The tidal bias parameter $\gamma_2$ was fixed to the excursion set relation of \citet{SheChaSco1304,EggScoCro2011} (and thus fully determined by the value of $b_1$) and all other parameters not appearing in Table~\ref{tab:validation_params} have been fixed to zero. The number densities given here 
enter in the computation of the associated covariance matrices.

%%%%%%%%%%%%%%%%%%%%%%%%%%%%%%%%%%%%%%%%%%%%%%%%%%

% Don't change these lines
\bsp	% typesetting comment
\label{lastpage}
\end{document}